\documentclass[reprint,pra,aps]{revtex4-2}
\pdfoutput=1
\usepackage{amsmath,amssymb}
\usepackage{graphicx}
\usepackage{color}
\usepackage{dcolumn}
\usepackage{bm}
\usepackage{array}
\usepackage{multirow}
\usepackage{hyperref}
\hypersetup{colorlinks=true,linkcolor=blue,citecolor=red}
\usepackage{slashbox}
\usepackage{calc}

\begin{document}
\title{Instability of multi-mode systems with quadratic Hamiltonians}

\author{Xuanloc Leu}
\author{Xuan-Hoai Thi Nguyen}
\author{Jinhyoung Lee}
\email{hyoung@hanyang.ac.kr}
\affiliation{Department of Physics, Hanyang University, Seoul 04763, Korea}

\received{\today}

\begin{abstract}
We present a novel geometric approach for determining the unique structure of a Hamiltonian and establishing an instability criterion for quantum quadratic systems. Our geometric criterion provides insights into the underlying geometric perspective of instability: A quantum quadratic system is dynamically unstable if and only if its Hamiltonian is hyperbolic. By applying our geometric method, we analyze the stability of two-mode and three-mode optomechanical systems. Remarkably, our approach demonstrates that these systems can be stabilized over a wider range of system parameters compared to the conventional rotating wave approximation (RWA) assumption. Furthermore, we reveal that the systems transit their phases from stable to unstable, when the system parameters cross specific critical boundaries. The results imply the presence of multistability in the optomechanical systems.

\end{abstract} 

\maketitle

\newcommand{\bra}[1]{\left<#1\right|}
\newcommand{\ket}[1]{\left|#1\right>}
\newcommand{\abs}[1]{\left|#1\right|}
\newcommand{\expt}[1]{\left<#1\right>}
\newcommand{\braket}[2]{\left<{#1}|{#2}\right>}
\newcommand{\commt}[2]{\left[{#1},{#2}\right]}

\newcommand{\tr}[1]{\mbox{Tr}{#1}}

\newcommand{\new}[1]{\textcolor{blue}{#1}}
\newcommand{\issue}[1]{\textcolor{red}{#1}}
\newcommand{\I}{\mathsf{i}}
\newcommand{\E}{\mathsf{e}}

\label{sec:exactsol}

\section{Introduction}

Quantum information processing and communication (QIPC) has developed rapidly in the past few decades. Optomechanical systems, composite systems of light and mechanical modes interacting by radiation-pressure force, have been proposed for QIPC applications~\cite{App,MTK,Rev2}. The theoretical and experimental studies on effects by radiation pressure on large objects were conducted in the context of interferometers~\cite{Braginsky}. The optical bistability in a Fabry-Perot resonator was experimentally demonstrated to vary by radiation pressure~\cite{Dorsel}. Various quantum phenomena effected by the optomechanical interaction have been observed, including squeezed optomechanics~\cite{Squeezing}, cooling of the mechanical mode to its motional ground state~\cite{Cooling}, generation of optomechanical entanglement~\cite{Entanglement}, optomechanical induced transparency~\cite{OMIT}, and optomechanical transduction~\cite{Transducer}. As a prerequisite for QIPC applications, the stability of optomechanical systems has received attention~\cite{Squeezing,Cooling,Entanglement,OMIT,Transducer,Stability,Amplifier,o3}.

Most studies on the optomechanical systems have focused on stable regions, and the physical mechanism of instability remains unraveled in more general ranges of physical parameters. The stability condition was investigated, when the optical cavity is driven by a coherent field; the strength of driving field determines the effective coupling between the cavity and mechanical modes \cite{MTK}. In particular, it has been studied in the assumption of the rotating wave approximation (RWA), with the red-sideband laser ($\omega_{L} + \Omega \approx \omega_{\text{cav}}$, where $\omega_L$, $\Omega$, and $\omega_{\text{cav}}$ are laser, mechanical, and cavity frequencies, respectively) and/or blue-sideband laser ($\omega_{L} \approx \omega_{\text{cav}} + \Omega$)~\cite{Squeezing,Cooling,Entanglement,OMIT,Transducer,Stability,Amplifier,o3}. In RWA, the red-detuned laser stabilizes a two-mode optomechanical system, while the blue-detuned laser renders it unstable~\cite{Amplifier}. For a three-mode optomechanical system, the power of red-detuned laser needs to be stronger than that of blue-detuned laser to stabilize in RWA~\cite{o3}. In this paper we generalize and extend the stability analysis of optomechanical systems to the full range of detuning frequencies and effective optomechanical couplings, going beyond RWA. We also explore the physical mechanism of instability.

Methods of stability analysis on a quantum system include Routh-Hurwitz method~\cite{Hur,Rou}. This method can be applied even to a system of nonlinear differential equations. It requires, on the other hand, sophisticated tools to determine stability conditions, and hard to unravel the underlying physics on the stability~\cite{RH}. An alternative method can be employed when a system's Hamiltonian is given in a quadratic form with constant coefficients. It has been applied well to a classical system~\cite{Meyer}. In this work we extend the method to a quantum quadratic system~\cite{Kus} by employing a geometric picture. The geometric picture is motivated by the observation:  A single-mode orbit diverges on phase space if it is governed by a hyperbolic equation. We generalize the geometric picture for multi modes and find a geometrically unique structure of Hamiltonian for a time-independent quantum quadratic system.

In this paper, we derive an instability criterion for a quantum quadratic system with constant coefficients by the geometric approach (Sec.~\ref{sec2}). The criterion reveals the underlying geometric perspective of instability: A time-independent quantum quadratic system is dynamically unstable if and only if its Hamiltonian can transform to be hyperbolic. We then apply the geometric method to the stability analysis of two-mode and three-mode optomechanical systems (Sec.~\ref{sec3}). Expanding the range of physical parameters beyond RWA, we show that the two-mode system can be controlled dynamically stable even in the blue-detuning regime, and the three-mode system can be stabilized irrespective of the relative power of the red-detuned laser to the blue-detuned. We also show that the system transits its phase from stable to unstable, as the system parameters cross the critical boundaries, i.e., the parameter values at which the Hamiltonian changes its geometric type from circular to hyperbolic. Additionally, we show that the Hamiltonian of an $N$-mode unstable quantum quadratic system is transformed to the hyperbolic Hamiltonian by some unitary operation followed by some symplectic operation in Appendix \ref{appA}, giving the explicit forms for a two-mode (three-mode) optomechanical system in Appendix \ref{appB} (\ref{appC}). 

\section{Hyperbolic Hamiltonians and unstable quadratic systems}\label{sec2}

We start with the observation that a single-mode orbit diverges on phase space of position $x$ and momentum $p$ if it is governed by a hyperbolic equation, i.e., $\alpha p^{2} - \beta x^2 = \text{const}$ with $\alpha  >0$ and $\beta \geq 0$~\cite{Jose}. If $\beta < 0$, the orbit becomes elliptic and is bounded in a finite region on phase space. From this observation, we employ a geometric approach of orbits on phase space with Hamiltonian diagonalized under symplectic transformations, so that we analyze the instability of a quadratic system. Here, the diagonalization means that Hamiltonian is given by squares of canonical variables with constant coefficients, or equivalently, its equation-of-motion matrix is diagonalized.

To introduce the principal idea for a quantum system, we consider the simplest case of a single-mode {\em unstable} system, whose Hamiltonian is given as
\begin{equation}\label{hy}
\hat{H} = \alpha \hat{p}^{2} - \beta \hat{x}^{2},
\end{equation}
where $\alpha$, $\beta$ are the positive real coefficients, $\hat{p}$ and $\hat{x}$ the momentum and position operators, respectively. The observable operators satisfy the canonical commutation relations $[\hat x,\hat p]=i $ in unit of $\hbar = 1$. The solution to the Heisenberg equation, governed by the Hamiltonian in Eq.~\eqref{hy}, is given by
\begin{eqnarray}
 \hat{x}(t) &=& \hat{x}(0)  \cosh{(2t\sqrt{\alpha \beta})} +  \hat{p}(0) \sinh{(2t \sqrt{\alpha \beta})}, \\
 \hat{p}(t) &=& \hat{p}(0) \cosh{(2t \sqrt{\alpha \beta})}  +  \hat{x}(0) \sinh{(2t \sqrt{\alpha \beta})}, 
 \end{eqnarray}
that grow indefinitely as a function of time. In other words, the orbit diverges on the phase space. The divergence (or the instability) originates from the hyperbolic structure of Hamiltonian in Eq.~\eqref{hy}. The Hamiltonian in the hyperbolic form of Eq.~\eqref{hy} is said {\em hyperbolic}. That the Hamiltonian is hyperbolic is a sufficient condition for the system to be unstable. 
 
We prove that it is also a necessary condition when the Hamiltonian is quadratic, in other words, an unstable single-mode system is governed by a hyperbolic Hamiltonian. The most general form of a single-mode quadratic Hamiltonian is given by
\begin{eqnarray}\label{Ha1}
\hat{H}_{g}  = \frac{1}{2} \bm{\hat{\xi}}^T \bm{V} \bm{\hat{\xi}} = 
\begin{pmatrix}
\hat{x} & \hat{p}
\end{pmatrix}
\begin{pmatrix}
\beta_1 & \gamma_1 \\
\gamma_1 & \alpha_1
\end{pmatrix}
\begin{pmatrix}
\hat{x} \\
\hat{p}
\end{pmatrix},
\end{eqnarray}
where $\hat{\bm{\xi}}=(\hat{x},\hat{p})^T$, and $\alpha_{1}$, $\beta_{1}$ and $\gamma_1$ are real numbers.
On one hand, the stability analysis method in Ref.~\cite{Meyer} shows that the system is unstable if and only if
\begin{equation}\label{con1}
\alpha_{1} \beta_{1} \leq \gamma_{1}^{2}.
\end{equation}
On the other hand, we apply some symplectic transformation $\bm{S}$ to diagonalize $\bm{V}$ in Eq.~\eqref{Ha1}, as in Ref.~\cite{Arv}, into a {\em geometric} form of
\begin{equation}\label{Ha2}
\hat{H}_{g} =  \frac{1}{2} \hat{\bm{\Xi}}^T \bm{D} \hat{\bm{\Xi}} = 
\begin{pmatrix}
\hat{X} & \hat{P}
\end{pmatrix}
\begin{pmatrix}
\beta' & 0 \\
0 & \alpha'
\end{pmatrix}
\begin{pmatrix}
\hat{X} \\
\hat{P}
\end{pmatrix}
= \alpha' \hat{P}^{2} + \beta' \hat{X}^{2},
\end{equation}
where $\hat{\bm \Xi} = \bm{S} \hat{\bm \xi}$, ${\bm J} {\bm D} = \bm{S} \bm{J V} \bm{S}^{-1}$, $\bm{J}$ is a skew symmetric matrix with elements $J_{jk} = -i [\hat{\xi}_j, \hat{\xi}_k] = -i [\hat{\Xi}_j, \hat{\Xi}_k]$, $\alpha' = [(\alpha_{1}+\beta_{1}) + \sqrt{(\alpha_{1}-\beta_{1})^{2} + 4 \gamma_{1}^{2}}]/2$, and $\beta' = [(\alpha_{1}+\beta_{1}) - \sqrt{(\alpha_{1}-\beta_{1})^{2} + 4 \gamma_{1}^{2}}]/2$. Here,  $\hat{\bm \Xi} = (\hat{X}, \hat{P})^T$ are the new canonical variables transformed from the original $\hat{\bm \xi} = (\hat{x}, \hat{p})^T$ by the symplectic matrix
\begin{equation}
\label{eq:dea}
\bm{S} =
\begin{pmatrix}
\sin{\theta} & \cos{\theta} \\
- \cos{\theta} & \sin{\theta}
\end{pmatrix},
\end{equation}
where $\tan 2\theta = 2\gamma_{1} /(\alpha_{1}-\beta_{1})$. 
The geometric form of Hamiltonian $\hat{H}_g$ in Eq.~\eqref{Ha2} is hyperbolic with $\beta' \le 0$, if $\alpha_{1} \beta_{1} \le \gamma_{1}^2$. This condition coincides with the one in Eq.~\eqref{con1} that the system is unstable.  It is seen that the hyperbolic Hamiltonian is the necessary condition for the system to be unstable, as well as sufficient. A special case is considered that $\alpha_{1} \beta_{1} = \gamma_{1}^2$, for which $\beta' = 0$ and $\alpha' = \alpha_{1} + \beta_{1} $. In the case, the system is `free' with $\hat{H}_{g} =  \alpha' \hat{P}^{2}$, which we also call `lineal' in the geometric perspective. We say the lineal (or free) Hamiltonian belongs to the hyperbolic, for a sake of simplicity, as its orbit also is unbounded on the phase space. It is remarkable that the hyperbolic Hamiltonian is the necessary and sufficient condition for the instability of a single-mode quadratic system. 

We generalize the equivalence between the hyperbolic Hamiltonian and the instability, for $N$-mode quadratic systems. Let us consider Hamiltonian given in general by
\begin{eqnarray}\label{quad-H}
\hat{H}_{N} &=&\sum_{j,k = 1}^N \left(\alpha_{jk}\hat{p}_{j} \hat{p}_{k}  + \beta_{jk} \hat{x}_{j} \hat{x}_{k} + \gamma_{jk} \hat{x}_{j} \hat{p}_{k} + \gamma_{jk} \hat{p}_{k} \hat{x}_{j}\right)\nonumber\\
&=& \frac{1}{2} \hat{\bm{\xi}}^T \bm{V} \hat{\bm{\xi}},
\end{eqnarray}
where $\alpha_{jk}, \beta_{jk} $, $\gamma_{jk}$ are time-independent real coefficients with $\alpha_{jk} = \alpha_{kj}$, $\beta_{jk} = \beta_{kj}$, $\hat p_j$ and $\hat x_j$ are of mode $j = 1, ..., N$, and $\hat{\bm{\xi}} = (\hat{x}_1, ..., \hat{x}_N, \hat{p}_1, ..., \hat{p}_N)^T$. To this end, we transform $\bm{V}$ in Eq.~\eqref{quad-H} by some symplectic matrix $\bm{S}$ and take an interaction picture by some unitary transformation $\hat{U}_I(t)$, so as to obtain a geometric form of Hamiltonian $\hat{H}_G$. By the geometric Hamiltonian we investigate the instability of $N$-mode quadratic system as testing whether it contains any modes of local hyperbolic Hamiltonians. 

Quadratic Hamiltonians in the form of Eq.~\eqref{quad-H} can be classified into 6 types of Jordan normal forms with respect to eigenvalues of $\bm{J V}$ (see Append. \ref{appA} and Refs.~\cite{Kus, LM}), where $\bm{J}$ is a skew symmetric matrix for $N$ modes with elements $J_{jk} = - i [\hat{\xi}_j, \hat{\xi}_k]$. We show in Appendix \ref{appA} that, for all the types, Hamiltonian~\eqref{quad-H} is rewritten by
\begin{equation}\label{dig}
\hat{H}_{N}=\hat{H}_G+\hat{H}_I,
\end{equation}  
where $\hat{H}_G$ is a geometric Hamiltonian decomposed into a sum of modal geometric Hamiltonians,
\begin{equation}\label{Geo-H}
\hat{H}_G=\sum_{k=1}^N\hat{H}_k=\sum_{k=1}^N\left(\alpha'_{k}\hat{P}^2_k + \beta'_{k}\hat{X}^2_k\right),
\end{equation}
where $\hat{H}_k$ is the geometric Hamiltonian of mode $k$,  $\alpha'_{k}$ and $\beta'_{k}$ are its real coefficients, and new canonical variables $\hat{P}_k$ and $\hat{X}_k$ result from $\hat{p}_k$ and $\hat{x}_k$ by the symplectic transformation $\bm{S}$. The form of interaction Hamiltonian $\hat{H}_I$ varies, depending on the type, while $\hat{H}_I$ always commutes with $\hat{H}_G$, i.e., $[\hat{H}_I, \hat{H}_G] = 0$, as in Appendix \ref{appA}. The commutation leads us to take the interaction picture by unitary transformation $\hat{U}_I(t) = \exp(-i t \hat{H}_I)$, so that the Hamiltonian in the picture is given by 
\begin{equation}\label{IP-H}
\hat{H}_{G}(t) = \hat{U}_I(t) \hat{H}_{G} \hat{U}^\dag_I(t) = \hat{H}_G.
\end{equation}
It is worth noting that the Hamiltonian $\hat{H}_{G}(t) = \hat{H}_G$ in the interaction picture, i.e., independent of time, thanks to the commutation of $[\hat{H}_I, \hat{H}_G]=0$. Thus, the general quadratic system is governed in the interaction picture by the geometric Hamiltonian $\hat{H}_G$ in Eq.~\eqref{Geo-H}, consisting of independent modes, regardless of its instability. Looking further into the geometric structure of Hamiltonian \eqref{Geo-H}, we see that $\hat{H}_G$ contains at least one mode of $\hat{H}_k$ hyperbolic when the system is dynamically unstable (see Appendix \ref{appA}). For a sake of simplicity, we say a multi-mode Hamiltonian is hyperbolic when it contains at least one mode of hyperbolic Hamiltonian. 

The equivalence between the instability and the hyperbolic form of Hamiltonian induces an instability criterion: {\em A (time-independent) quantum quadratic system is dynamically unstable if and only if its (transformed) Hamiltonian is hyperbolic,} which is one of our main results. This criterion allows us to analyze the instability in terms of the geometric Hamiltonian and also provides us the physical insight for stabilizing quantum systems.

\section{Instability of optomechanical systems}\label{sec3}

Our geometric method is applied to dynamical stability of an optomechanical system. In one case of two modes (Sec. \ref{2mode}), a mechanical oscillator of one mode is a mirror to build an optical cavity of the other mode, as in Fig.~\ref{fig:OMS}. In the other case (Sec. \ref{3mode}), one mode is a middle mirror separating two optical cavities of the other two modes, respectively, as in Fig.~\ref{fig:4}. The optomechanical system has been studied most at sideband driving frequencies, where it is stabilized by the red-detuned pumping laser, whereas it becomes unstable by the blue-detuned laser~\cite{Amplifier}. These opposite effects of the red-detuned and blue-detuned lasers can cooperate as they simultaneously affect a three-mode optomechanical system. For a three-mode optomechanical system, thus, one may require that the power of the red-detuned laser is stronger than that of the blue-detuned laser~\cite{o3}. We show that these constraints are relaxed beyond the sideband interaction limit.

\subsection{Two modes}\label{2mode}
\begin{figure}[h!]
\begin{center}
\includegraphics[height=0.15\textwidth]{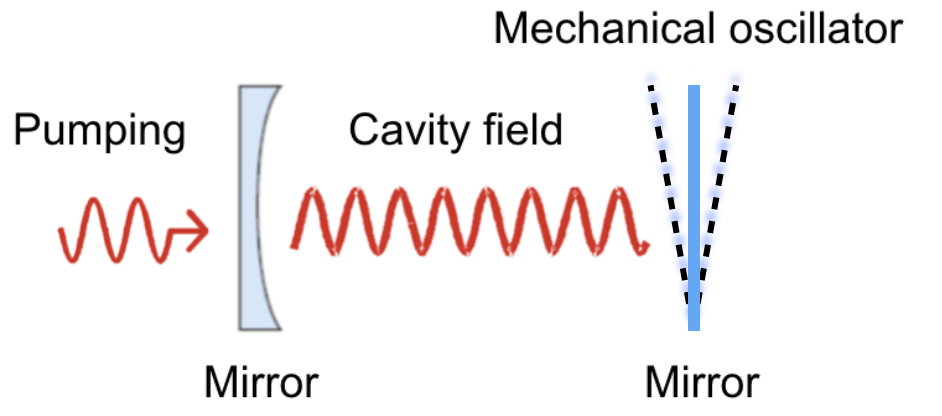}
\end{center}
\caption{Schematic of a two-mode optomachanical system.}
\label{fig:OMS}
\end{figure}

We consider an optical Fabry-Perot cavity consisting of one mirror firmly fixed and the other movable, as depicted in Fig.~\ref{fig:OMS}. The movable mirror is modeled as a quantum harmonic oscillator $b$ with annihilation operator $\hat{b}$ and frequency $\Omega$. As it is affected by the radiation pressure of cavity, the mechanical oscillator $b$ is coupled to the optical cavity $a$ with annihilation operator $\hat{a}$ and frequency $\omega_{\text{cav}}$. The cavity is pumped by a laser field with strength $\kappa_\text{in}$ and frequency $\omega_L$. The Hamiltonian in the rotating frame of the laser frequency is given~\cite{MTK} by
\begin{equation}\label{H2m}
\hat{H}_{\text{cm}} =  \Delta' \, \hat{a}^{\dag}  \hat{a}  +  \Omega \, \hat{b}^{\dag}  \hat{b} - \kappa_{0}  \hat{a}^{\dag} \hat{a}(  \hat{b} + \hat{b}^{\dag}) -  (\kappa_{\text{in}}\hat{a}^{\dag} + \kappa^*_{\text{in}}\hat{a}),
\end{equation}
 where $\Delta' = \omega_{\text{cav}} - \omega_L $ is the detuning between cavity mode and pumping laser, and $\kappa_0$ is the optomechanical coupling constant. We "linearize" Hamiltonian in Eq.~\eqref{H2m} by assuming the limit of strong pumping, $|\alpha_{s}|^2 \gg \langle \delta \hat{a}^\dag \delta \hat{a} \rangle$, where $\alpha_{s}$ is the steady-state amplitude and the deviation $\delta\hat{a} = \hat{a} - \alpha_{s}$. Similarly, we consider the deviation $\delta \hat{b} = \hat{b} - \beta_s$ from the steady-state amplitude $\beta_s$ for the mechanical mode. The approximated Hamiltonian up to the second order of the deviations is given in a quadratic form of two modes by
 \begin{equation}\label{H2m-lin}
\hat{H}_{\text{cm-lin}} = \Delta \delta \hat{a}^{\dag}  \delta \hat{a} +  \Omega  \delta \hat{b}^{\dag} \delta \hat{b} + \left( \kappa \delta \hat{a}^\dagger+\ \kappa^{*} \delta\hat{a}\right)(\delta \hat{b}^\dagger + \delta \hat{b}),
\end{equation} 
where Lamb-shifted detuning $\Delta = \Delta' - \kappa_0 (\beta_s + \beta_s^*) = \Delta' - 2\kappa_0^2|\alpha_s|^2 / \Omega$, and effective coupling constant $\kappa=-\kappa_0\alpha_{s}=-\kappa_0\kappa_{\text{in}}/\Delta$. Here the amplitudes of stead state are given by $\alpha_s =\kappa_{\text{in}} / \Delta$ and $\beta_s = \kappa_0|\alpha_s|^2 / \Omega$~\cite{MTK}. The vacuum states of the deviation modes $\delta a$ and $\delta b$ are given in terms of the original modes $a$ and $b$ by 
 \begin{equation}\label{vacuadm}
\ket{0}_{\delta a} = \ket{\alpha_s}_a \quad \text{and} \ket{0}_{\delta b} = \ket{\beta_s}_b,
\end{equation} 
where $\ket{\alpha_s}$ and $\ket{\beta_s}$ are coherent states of modes $a$ and $b$.

The linearized Hamiltonian $\hat{H}_\text{cm-lin}$ derived above governs the quantum dynamics of the system near a given steady-state point $\{\alpha_s,\beta_s\}$. The analysis of stability against small fluctuations near the steady state is based on the equations of motion for the deviation modes  (associated with $\hat{H}_\text{cm-lin}$) in the presence of noises~\cite{Squeezing,Cooling,Entanglement,OMIT,Transducer,Stability,Amplifier,o3}. The requirement for stability can then be derived by applying, e.g., the Routh-Hurwitz criterion~\citep{RH}. However, the analytic expressions are quite cumbersome and they can just be practical in numerical works, in finding the threshold values for a given set of system parameters. In this work, adopting the geometric approach, we explore the mechanism of instability that arises from the optomechanical interaction, neglecting the noise effects, and we analyze the quantum dynamics of fluctuations from the linearized Hamiltonian $\hat{H}_\text{cm-lin}$.

To apply our geometric method of instability, we rewrite Hamiltonian~\eqref{H2m-lin} in terms of quadratures,
\begin{eqnarray}
\label{H2m-q}
\hat{H}_\text{cm-lin} 
&=&\frac{\Delta}{2}(\hat{p}^2_1+\hat{x}^2_1)+\frac{\Omega}{2}(\hat{p}^2_2+\hat{x}^2_2)+ 2(\kappa_r\hat{x}_1+\kappa_i\hat{p}_1)\hat{x}_2\nonumber\\ 
&=& \frac{1}{2} \hat{\bm{\xi}}^T \bm{V} \hat{\bm{\xi}},
\end{eqnarray}
where $\hat{x}_1=(\delta\hat{a}^\dagger+\delta\hat{a})/\sqrt{2}$, $\hat{p}_1=i(\delta\hat{a}^\dagger-\delta\hat{a})/\sqrt{2}$, $\hat{x}_2=(\delta\hat{b}^\dagger+\delta\hat{b})/\sqrt{2}$, $\hat{p}_2=i(\delta\hat{b}^\dagger-\delta\hat{b})/\sqrt{2}$, $\kappa_r = \text{Re}(\kappa)$, and $\kappa_i =\text{Im}(\kappa)$. It is reminded that $\hat{\bm{\xi}}=(\hat{x}_1,\hat{x}_2,\hat{p}_1,\hat{p}_2)^T$. We then find the geometric Hamiltonian $\hat{H}_G$ from the original Hamiltonian $\hat{H}_\text{cm-lin}$ in Eq.~\eqref{H2m-q}, as in Sec.~\ref{sec2} for $N=2$ (see Appendix \ref{appB}).

\begin{table}
\begin{center}
\caption{Geometric kinds of modal Hamiltonians $\hat{H}_{1,2}$ and stability for two-mode optomechanical system with respect to critical parameters $K_{R} = \sqrt{\Omega |\Delta|/4}$ and $K_B = \sqrt{( \Delta^{2} -  \Omega^{2})^{2}/16 \Omega |\Delta |}$. Cases (a)-(c) are in the red-detuning regime (detuning $\Delta > 0$), cases (e)-(g) in the blue-detuning regime ($\Delta < 0$), and the other case (d) of resonance ($\Delta = 0$). A modal Hamiltonian $\hat{H}_k$ is of a geometric kind either circular, hyperbolic, or lineal (free), when it is in a form of $\hat{H}_\text{circular} = \lambda (\hat{P}^{2} + \hat{X}^{2})$, $\hat{H}_\text{hyperbolic} =  \lambda (\hat{P}^{2} - \hat{X}^{2})$, or $\hat{H}_\text{lineal} =  \lambda \hat{P}^{2}$ for some real number $\lambda$, whose explicit expression is given in Appendix \ref{appB}.}
 \label{tab:1}
 \begin{ruledtabular}
\begin{tabular}{lccc} 
Case &  $\hat{H}_1$ & $\hat{H}_2$ & Stability \\ 
\hline
(a) $\Delta > 0$, $ |\kappa| < K_R $   & circular & circular & stable \\ 
(b) $\Delta > 0$, $ |\kappa| = K_R $   & circular & lineal & unstable \\ 
(c) $\Delta > 0$, $ |\kappa| > K_R $   & circular & hyperbolic & unstable \\ 
\hline
(d) $\Delta = 0$  & circular & lineal & unstable \\
\hline
(e) $\Delta < 0$, $ |\kappa| < K_B $  & circular & circular & stable \\ 
(f) $\Delta < 0$, $ |\kappa| = K_B $   & lineal & lineal  & unstable \\ 
(g) $\Delta < 0$, $ |\kappa| > K_B $   & hyperbolic & hyperbolic & unstable \\ 
\end{tabular}
\end{ruledtabular}
\end{center}
\end{table}

Table~\ref{tab:1} summaries 7 cases of geometric Hamiltonians $\hat{H}_G$ and their stabilities in terms of parameters $\Delta, K_R = \sqrt{\Omega |\Delta|/4},$ and $K_B=\sqrt{(\Delta^2 - \Omega^2)^2 /16 \Omega |\Delta|}$, where three cases (a)-(c) are in red-detuning regime with detuning $\Delta > 0$, other three cases (e)-(g) are in blue-detuning regime with $\Delta < 0$, and the other case (d) is of resonance $\Delta = 0$. Here each modal Hamiltonian $\hat{H}_{k}$ is said either circular, hyperbolic, or lineal (or free), when it is in a form of $\hat{H}_\text{circular} = \lambda (\hat{P}^{2} + \hat{X}^{2})$, $\hat{H}_\text{hyperbolic} =  \lambda (\hat{P}^{2} - \hat{X}^{2})$, or $\hat{H}_\text{lineal} =  \lambda \hat{P}^{2}$ for some real number $\lambda$, whose explicit expression is given in Appendix \ref{appB}. 
The system is stable in case (a) or (e) as its geometric Hamiltonian $\hat{H}_G$ consists of two circular modal Hamiltonians, whereas the other cases are unstable as $\hat{H}_G$ contains at least one hyperbolic (or lineal) modal Hamiltonian(s). The stability conditions are given in case (a) by $\Delta > 0$ and $\Omega \Delta - 4 |\kappa|^2 > 0$, and in case (e) by $\Delta < 0$ and $(\Delta^2 + \Omega^2)^2 >4\Omega \Delta (\Omega \Delta - 4|\kappa|^2)$. Both cases are merged to
\begin{equation}\label{StaCon}
(\Delta^{2} + \Omega^{2})^{2} > 4 \Omega \Delta  ( \Omega \Delta - 4 |\kappa|^{2}) > 0.
\end{equation}
We prove the inequalities~\eqref{StaCon} hold in both cases of (a) and (e): For $\Delta > 0$, the first inequality is trivial and the second inequality holds by the stability conditions in case (a). For $\Delta < 0$, the second inequality is trivial and the first inequality holds by the stability conditions in case (e). 
The stability conditions in Eq.~\eqref{StaCon} coincide with those in Ref.~\cite{Meyer}.

To explain the origin of hyperbolic Hamiltonian, we transform the linearized Hamiltonian \eqref{H2m-lin} to, in the interaction picture,
\begin{eqnarray}\label{H2m-int}
\hat{H}_{\text{cm-int}}(t) 
&=&  \hat{U}_0^\dagger(t)\hat{H}_\text{cm-lin}\hat{U}_0(t)-i\hat{U}_0^\dagger(t)\frac{d \hat{U}_0(t)}{dt}\nonumber\\
&=&\hat{H}_{\text{bs}}(t) + \hat{H}_{\text{sq}}(t),
\end{eqnarray}
where $\hat{U}_0(t)=\exp[-it(\Delta\delta\hat{a}^\dagger\delta\hat{a}+\Omega\delta\hat{b}^\dagger\delta\hat{b})]$, and
\begin{eqnarray}
\hat{H}_\text{bs}(t)&=&\kappa \delta\hat{a}^\dagger\delta\hat{b}e^{it(\Delta-\Omega)}+\kappa^*\delta\hat{a}\delta\hat{b}^\dagger e^{-it(\Delta-\Omega)},\\
\hat{H}_\text{sq}(t)&=&\kappa \delta\hat{a}^\dagger\delta\hat{b}^\dagger e^{it(\Delta+\Omega)}+\kappa^* \delta\hat{a}\delta\hat{b}e^{-it(\Delta+\Omega)}.
\end{eqnarray}
The beam-splitter term $\hat{H}_\text{bs}$, which is resonant at $\Delta=\Omega$, describes the exchange of excitation between the optical and mechanical modes, where excitation is created in one mode while being destroyed in the other. On the other hand, the two-mode squeezing term $\hat{H}_\text{sq}$, which is resonant at $\Delta=-\Omega$, represents the down-conversion process that generates or destroys excitations in both modes simultaneously. While the beam-splitter term $\hat{H}_\text{bs}$ can cause an exchange of energy between the modes, the two-mode squeezing term $\hat{H}_\text{sq}$ can lead to an unbounded increase in energy for both modes, potentially triggering a dynamical instability in the system. Indeed, using the geometric approach, we find that the geometric structure of $\hat{H}_\text{bs}$ is circular for all values of $\Delta$, $\Omega$, and $\kappa$, indicating that the system governed by $\hat{H}_\text{bs}$ is always stable. Whereas, the geometric structure of $\hat{H}_\text{sq}$ varies with the parameters, appearing circular when $2|\kappa|<|\Delta+\Omega|$, hyperbolic (lineal) when $2|\kappa|>|\Delta+\Omega|$ ($2|\kappa|=|\Delta+\Omega|$). That the hyperbolic Hamiltonian governs the system and causes the instability when $2|\kappa|>|\Delta+\Omega|$, is in line with our earlier assumption about the effect of the squeezing interaction $\hat{H}_\text{sq}$ on the energy of the system at resonance $\Delta=-\Omega$. This fact suggests that the squeezing interaction is the origin of the hyperbolic Hamiltonian. As a result, the system is dynamically unstable in a parameter region where the squeezing interaction contributes dominantly. We show in the following that this provides an intuitive interpretation for all cases in Table~\ref{tab:1}.

For $\Delta>0$, the counter-rotating (squeezing) term $\hat{H}_\text{sq}$ can be neglected by RWA under the condition of weak coupling $(|\kappa|\ll\Omega,\Delta)$ as $|\Delta-\Omega|\ll |\Delta+\Omega|$, so that the system is stable [case (a)]. When the coupling $|\kappa|$ increases and becomes comparable to $\Omega$ or/and $\Delta$, the effect of $\hat{H}_\text{sq}$ becomes significant, leading to instability in the system [cases (b) and (c)]. For $\Delta < 0$, we focus on the weak coupling ($|\kappa|\ll \Omega, |\Delta|$) even though our method also works in the strong coupling. In the week coupling, $H_\text{sq}$ is dominant rather than $H_\text{bs}$, as $| \Delta + \Omega | = | |\Delta| - \Omega | \ll | |\Delta| + \Omega | = | \Delta - \Omega |$. Now we take two extremely cases: near resonance of $||\Delta| - \Omega| \ll |\kappa|$ and far-off resonance of $||\Delta| - \Omega| \gg |\kappa|$. In the near resonance of $||\Delta| - \Omega| \ll |\kappa|$, it is clear that $H_\text{sq}$ dominates over $H_\text{bs}$, so that the system is unstable [cases (f) and (g)]. On the other hand, in the far-off resonance of $||\Delta| - \Omega| \gg |\kappa|$, both of $H_\text{sq}$ and $H_\text{bs}$ are small and it is necessary to employ the second-order perturbations \cite{Knight}. Then, the effective Hamiltonian for the dispersive far-off resonant interactions is given by
\begin{equation}
\label{Heff}
\hat{H}_\text{cm-int}(t)\approx\hat{H}_\text{eff}=\frac{2|\kappa|^2}{\Delta^2-\Omega^2}(\Omega\delta\hat{a}^\dagger \delta\hat{a}-\Delta\delta \hat{b}^\dagger \delta\hat{b}).
\end{equation}
When turning back to Schrodinger picture, the effective Hamiltonian is transformed to
\begin{equation}
\label{Sch-eff}
\hat{H}_\text{cm-lin}
\approx\bigg(\Delta+\frac{2\Omega|\kappa|^2}{\Delta^2-\Omega^2}\bigg)\delta\hat{a}^\dagger \delta\hat{a}
+\bigg(\Omega-\frac{2\Delta|\kappa|^2}{\Delta^2-\Omega^2}\bigg)\delta \hat{b}^\dagger \delta\hat{b}.
\end{equation}
Thus, in the far-off resonance, the effective Hamiltonian is that of free harmonic oscillators, so that the system is stable [case (e)]  .

\begin{figure}[t]
\begin{center}
\includegraphics[height=0.47\textwidth]{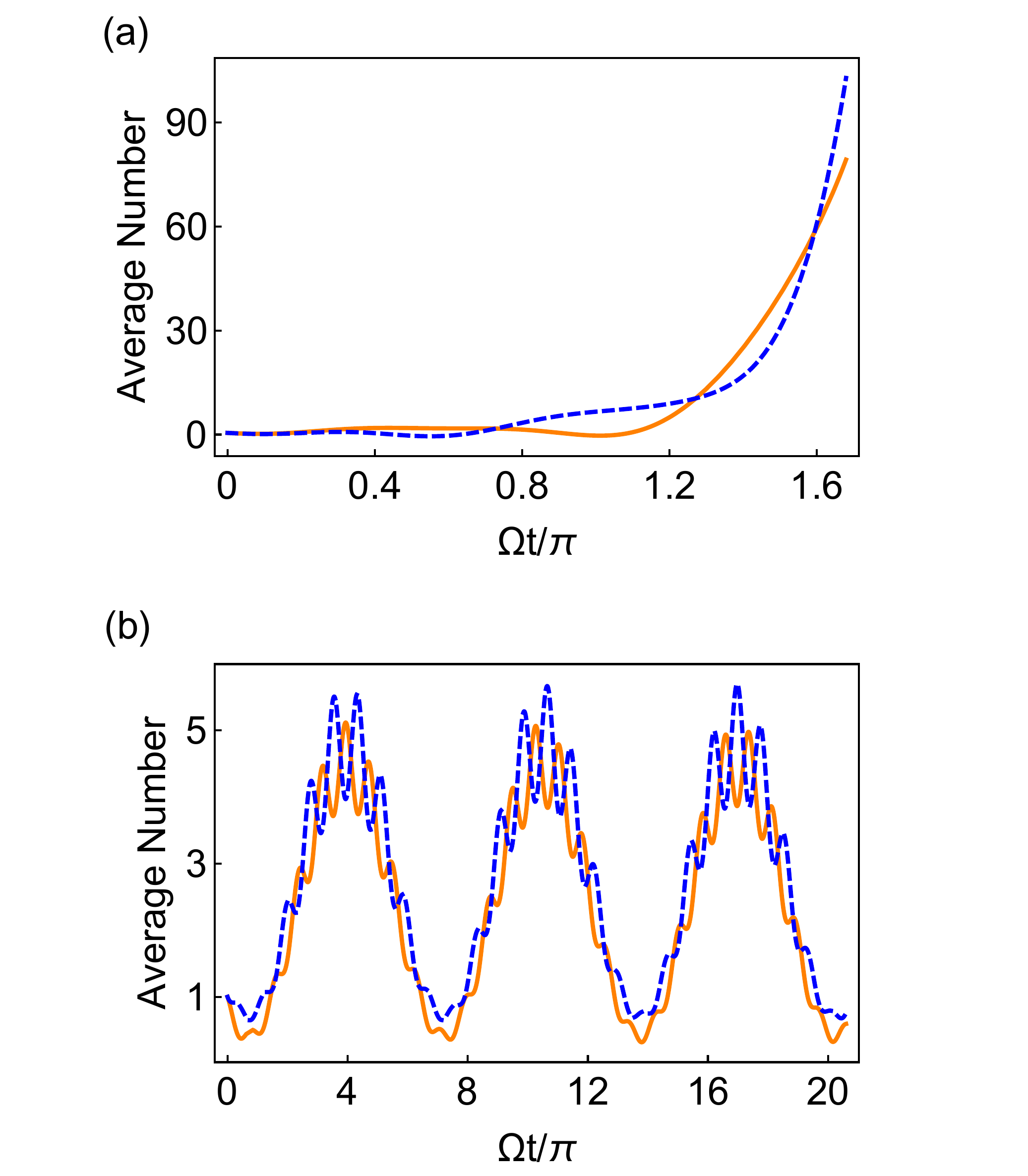}
\end{center}
\caption{Time evolution of average numbers, $\langle \delta \hat{a}^\dag \delta \hat{a} \rangle$ of the cavity mode (orange solid line) and $\langle \delta\hat{b}^\dag \delta\hat{b} \rangle$ of the mechanical mode (blue dashed line) in a two-mode optomechanical system with (a) $\Delta =1.5\, \Omega$ and $|\kappa| = 0.9\,\Omega$, and (b) $\Delta = -1.5\, \Omega$ and $|\kappa| = 0.2 \,\Omega$.}
\label{fig:2}
\end{figure}

We discuss two special situations of the red and blue sidebands by that the cavities are driven, included in Table~\ref{tab:1}. While the former belongs to case (a), the latter is part of case (g). Therefore, the known assumption that the system is stabilized by the red-detuned pumping laser while becoming unstable by the blue-detuned laser, is no longer valid in general. More precisely, the squeezing term of Hamiltonian triggers the instability in optomechanical systems. The effect is not solely determined by the type of detuning alone: The system can become unstable in the red-detuning regime [as shown in cases (b) and (c) and illustrated in Fig.~\ref{fig:2}(a)], whereas it can be controlled stable in the blue-detuning regime [case (e), Fig.~\ref{fig:2}(b)]. Fig.~\ref{fig:2}(a) shows that the average numbers diverge of the cavity and mechanical modes when the cavity is driven by a red-detuned laser with $\Delta=1.5\Omega>0$. In contrast, Fig.~\ref{fig:2}(b) demonstrates that the average numbers remain bounded in the blue-detuning regime with $\Delta=-1.5\Omega<0$.

\begin{figure}[t]
\begin{center}
\includegraphics[height=0.7\textwidth]{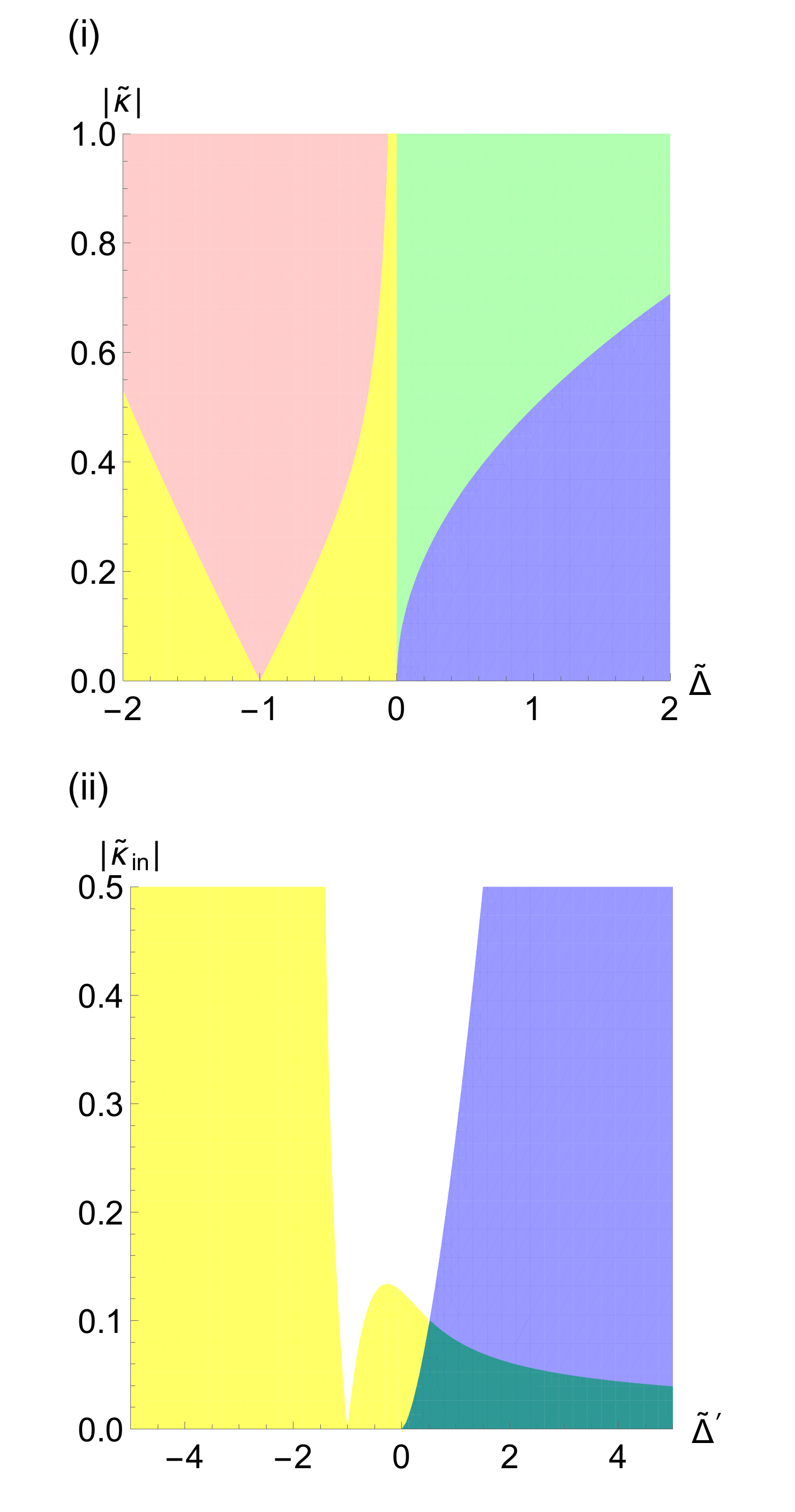}
\end{center}
\caption{(i) Stability diagram of a two-mode optomechanical system in terms of dimensionless parameters $\tilde{\Delta}:=\Delta/\Omega$ and $|\tilde{\kappa}|:=|\kappa|/\Omega$. The diagram represents all cases listed in Table~\ref{tab:1}: blue area: case (a), boundary between green and blue areas: case (b), green area: case (c), boundary between green and yellow areas: case (d), yellow area: case (e), boundary between yellow and pink areas: case (f), pink area: case (g). (ii) Stable areas corresponding to case (a) (blue) and case (e) (yellow) in terms of $\tilde{\Delta}^\prime:=\Delta^\prime/\Omega$ and $|\tilde{\kappa}_\text{in}|:=|\kappa_0\kappa_\text{in}|/\Omega^2$.\label{fig:3}}
\end{figure}

In addition, we present the diagrams of stability. Fig.~\ref{fig:3}(i) displays the stability diagram in terms of dimensionless parameters $\tilde{\Delta}:=\Delta/\Omega$ and $|\tilde{\kappa}|:=|\kappa|/\Omega$. This diagram encompasses all the cases described in Table~\ref{tab:1}. Particularly noteworthy are the critical boundaries representing cases (b), (d), and (f). These boundaries consist of the critical points $\{\Delta>0, |\kappa|=K_R\}$, $\{\Delta=0, \kappa\neq 0\}$, and $\{\Delta<0,|\kappa|= K_B\}$ for a given $\Omega$. At the critical points the Hamiltonian changes its geometric type from circular to hyperbolic. In other words, the system transits its phase from stable to unstable as the system parameters cross the critical boundaries. Consequently, the system can be controlled stable in both the red-detuning and blue-detuning regimes by adjusting the strength and frequency of the pumping field. We now focus our attention on the stable areas. In Fig.~\ref{fig:3}(ii), we depict a stability diagram illustrating cases (a) and (e) in terms of $\tilde{\Delta}^\prime:=\Delta^\prime/\Omega$ and $|\tilde{\kappa}_\text{in}|:=|\kappa_0\kappa_\text{in}|/\Omega^2$. The blue and yellow areas in Fig.~\ref{fig:3}(ii) correspond to two distinct stable steady states of the system, respectively, while the dark green area represents their overlap. These two stable steady states conform to the nonlinearity of the optomechanical interaction, where the system exhibits three possible steady states: two stable and one unstable~\cite{MTK}. Furthermore, both steady states found stable in the dark green overlapping region demonstrates the multi-stability characteristic of an optomechanical system~\citep{Bistable}. 

It is remarkable that we analyze the stability criteria of a two-mode optomechanical system by its geometric Hamiltonian in a wide range of system parameters. The extension beyond  the limitation of $\Delta=\pm \Omega$ enables us to explore the general stability condition, also to understand more insight into the mechanism causing the instability of the system. An unstable two-mode optomechanical system is governed by the hyperbolic Hamiltonian, which originates from the two-mode squeezing interaction term. Although focusing on the dynamics in the linearization approximation, the resulting geometric Hamiltonian provides an excellent background for a more general analysis, including noise and nonlinear coupling terms \citep{Meyer2,NEM}.

\subsection{Three modes}\label{3mode}

\begin{figure}[h!]
\begin{center}
\includegraphics[height=0.15\textwidth]{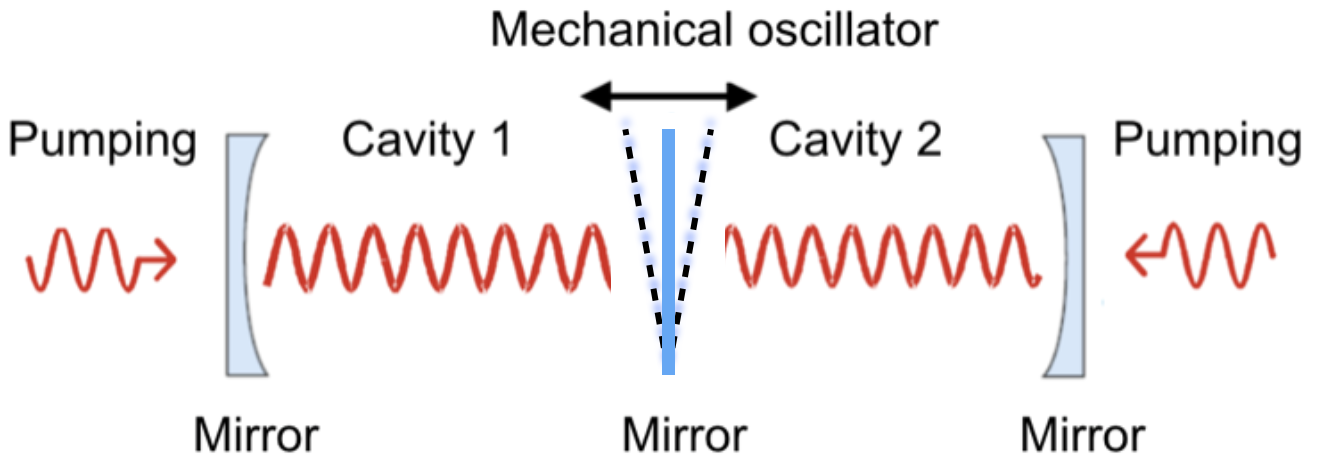}
\end{center}
\caption{Schematic of a three-mode optomachanical system.}
\label{fig:4}
\end{figure}

We consider a three-mode optomechanical system  consisting of two fixed mirrors and one movable mirror in between them, where the movable mirror is one mode of mechanical oscillator and two modes of optical cavities formed by the movable mirror and the two fixed mirrors, respectively, as in Fig.~\ref{fig:4}. The cavities are driven by two pump lasers, respectively. The Hamiltonian of system $\hat{H}_\text{cmc}$ is linearized, as done in Sec.~\ref{2mode},
\begin{eqnarray}\label{H3m-lin}
\hat{H}_{\text{cmc-lin}} 
&=& \sum_{j=1,2} \Delta_j  \delta \hat{a}_{j}^{\dag} \delta \hat{a}_{j} +  \Omega \delta\hat{b}^{\dag}  \delta\hat{b}\nonumber\\
&& + \sum_{j=1,2}  ( \kappa_{j}  \delta \hat{a}_{j}^{\dag} +  \kappa_{j}^{*} \delta \hat{a}_{j}) ( \delta\hat{b} +  \delta\hat{b}^{\dag}),
\end{eqnarray}
where $\delta\hat{a}_j$ ($\delta\hat{b}$) is the annihilation operator of deviation cavity mode $j$ (deviation mechanical mode), $\kappa_j$ the effective optomechanical coupling constant between cavity mode $j$ and the mechanical mode, and $\Delta_j$ the Lamb-shifted detuning between cavity mode $j$ and input laser $j$.

The stability condition and geometric kinds of modal Hamiltonian $\hat{H}_{1,2,3}$ for a general three-mode optomechanical system is shown in Appendix~\ref{appC}. Here, we focus on a simple case of $\Delta_1=s \Delta_2=\Delta$, where the sign $s=\pm$ ('$+$' means that both cavity modes are driven by either blue-detuned or red-detuned lasers, whereas '$-$' indicates that one cavity mode is driven by a red-detuned laser while the other is driven by a blue-detuned laser). The choice of $\Delta_1=\pm\Delta_2$ enables us to achieve a simple explicit expression for stability condition. Even so, it is general enough to go beyond RWA for any chosen value of $\Delta$ (rather than at the sideband interaction limit $|\Delta|=\Omega$). 

\begin{table*}
\begin{center}
\caption{Symplectic matrix $\bm{S}_c$, which transforms $(\delta\hat{a}_1,\delta\hat{a}_2,\delta\hat{a}_1^\dagger,\delta\hat{a}_2^\dagger)$ to $(\hat{A}_1,\hat{A}_2,\hat{A}_1^\dagger,\hat{A}_2^\dagger)$, for different values of $s$ and $\kappa_j$ with $j=1,2$.}
 \label{tab:2}
\begin{ruledtabular}
\begin{tabular}{lccc}
\multirow[c]{2}*{\backslashbox{$\bm{S}_c$}{$s, \kappa_j$}}& $s=+$ & \multicolumn{2}{c}{$s=-$} \\
\cline{3-4}
 &any $\kappa_j$& $|\kappa_1|>|\kappa_2|$ & $|\kappa_1|<|\kappa_2|$\\
\hline
 $\bm{S}_c$& $\frac{1}{\kappa_+}\begin{pmatrix}
\kappa_1^* & \kappa_2^* & 0 & 0 \\
\kappa_2 & -\kappa_1 & 0 & 0\\
0 & 0 & \kappa_1 & \kappa_2\\
0 & 0 & \kappa_2^* & -\kappa_1^*
\end{pmatrix}$ &
$\frac{1}{\kappa_-}\begin{pmatrix}
\kappa_1^* & 0 & 0 & \kappa_2 \\
0 & \kappa_1^* & \kappa_2 & 0\\
0 & \kappa_2^* & \kappa_1 & 0\\
\kappa_2^* & 0 & 0 & \kappa_1
\end{pmatrix}$ &
$\frac{1}{\kappa_-}\begin{pmatrix}
0 & \kappa_2^* & \kappa_1 & 0 \\
\kappa_2^* & 0 & 0 & \kappa_1\\
\kappa_1^* & 0 & 0 & \kappa_2\\
0 & \kappa_1^* & \kappa_2 & 0
\end{pmatrix}$\\
\end{tabular}
\end{ruledtabular}
\end{center}
\end{table*}

By symplectic transformation $\bm{S}_c$ (shown in Table~\ref{tab:2}), Hamiltonian $\hat{H}_\text{cmc-lin}$ \eqref{H3m-lin} is transformed to (except for the case $s=-1$ and $|\kappa_1|=|\kappa_2|$, which falls into case (5) of Table ~\ref{tab:4} in Appendix \ref{appC})
\begin{equation}\label{Hs}
\hat{H}_s=\hat{H}_{\hat{A}_1\text{-}\delta\hat{b}}+s \epsilon\Delta\hat{A}^\dagger_2\hat{A}_2.
\end{equation}
Here, the term $\hat{H}_{\hat{A}_1\text{-}\delta\hat{b}}$ describes a subsystem of two modes, $\hat{A}_1$ and $\delta\hat{b}$, coupling each other by the linearized optomechanical interaction
\begin{equation}\label{HA1b}
\hat{H}_{\hat{A}_1\text{-}\delta\hat{b}}=\epsilon\Delta\hat{A}^\dagger_1\hat{A}_1+\Omega\delta\hat{b}^\dagger\delta\hat{b}+\kappa_s(\hat{A}^\dagger_1+\hat{A}_1)(\delta\hat{b}^\dagger+\delta\hat{b}),
\end{equation} 
where $\kappa_s=\sqrt{\left||\kappa_1|^2+s|\kappa_2|^2\right|}$ and $\epsilon=$ sgn$\left[|\kappa_1|^2+s|\kappa_2|^2\right]$.
Because the harmonic oscillator mode $\hat{A}_2$ is stable and completely decouples from modes $\hat{A}_1$ and $\delta\hat{b}$, the analysis of stability relies on the two-mode optomechanical Hamiltonian $\hat{H}_{\hat{A}_1\text{-}\delta\hat{b}}$, enabling us to apply the results obtained in Sec.~\ref{2mode}. Accordingly, the stability condition for the system is 
\begin{equation}\label{sc-3m}
(\Delta^{2} + \Omega^{2})^{2} > 4 \Omega \Delta  ( \Omega \Delta - 4 \epsilon\kappa_s^2) > 0,
\end{equation}
which is the same as the condition in Eq.~\eqref{StaCon} with $\Delta$ and $|\kappa|$ replaced by $\epsilon\Delta$ and $\kappa_s$, respectively.
Geometric Hamiltonian of the sub-system including modes $\hat{A}_1$ and $\delta\hat{b}$ falls into one of the cases listed in Table~\ref{tab:1} conditional on the values of $s$, $\epsilon$, $\Delta$, and $\kappa_s$, as shown in Table~\ref{tab:3}. Mapping between the system parameters and cases (a)-(g) tells us specific kinds of modal geometric Hamiltonian for a given set of parameters. For instance, if $\epsilon \Delta<0$ and $\kappa_s>K_B$, the system is unstable in the manner of the case (g), where the geometric Hamiltonian of the system composing of two hyperbolic modal Hamiltonians associated with modes $\hat{A}_1$ and $\delta\hat{b}$ beside a circular Hamiltonian of mode $\hat{A}_2$.
\begin{table*}
\centering
\caption{Geometric Hamiltonian and stability of the subsystem governed by Hamiltonian $\hat{H}_{\hat{A}_1\text{-}\delta\hat{b}}$ in comparison with the cases from (a) to (g) described in Table \ref{tab:1}.}
 \label{tab:3}
\begin{ruledtabular}
\newlength\mylen
\setlength\mylen{(\widthof{$\epsilon\Delta>0$, any $s$}+58\tabcolsep)/3}
\begin{tabular}{l *{3}{|wc{\mylen}} *{3}{|wc{\mylen}}|c}
\multirow[c]{2}*{\backslashbox{Stability}{Cases}}&
\multicolumn{3}{c|}{$\epsilon\Delta>0$, any $s$}& 
\multicolumn{3}{c|}{$\epsilon\Delta<0$, any $s$}&
$\epsilon\Delta=0$,\\
\cline{2-7}
&
$\kappa_s<K_R$&$\kappa_s=K_R$&
$\kappa_s>K_R$& 
$\kappa_s<K_B$&$\kappa_s=K_B$&
$\kappa_s>K_B$&
any $s,\kappa_s$\\\hline
stable&
(a)&
&
& 
(e)&
&
&\\
unstable&
& 
(b)& 
(c)& 
&
(f)&
(g)&
(d)\\
\end{tabular}
\end{ruledtabular}
\end{table*}

\begin{figure}[t]
\begin{center}
\includegraphics[height=0.47\textwidth]{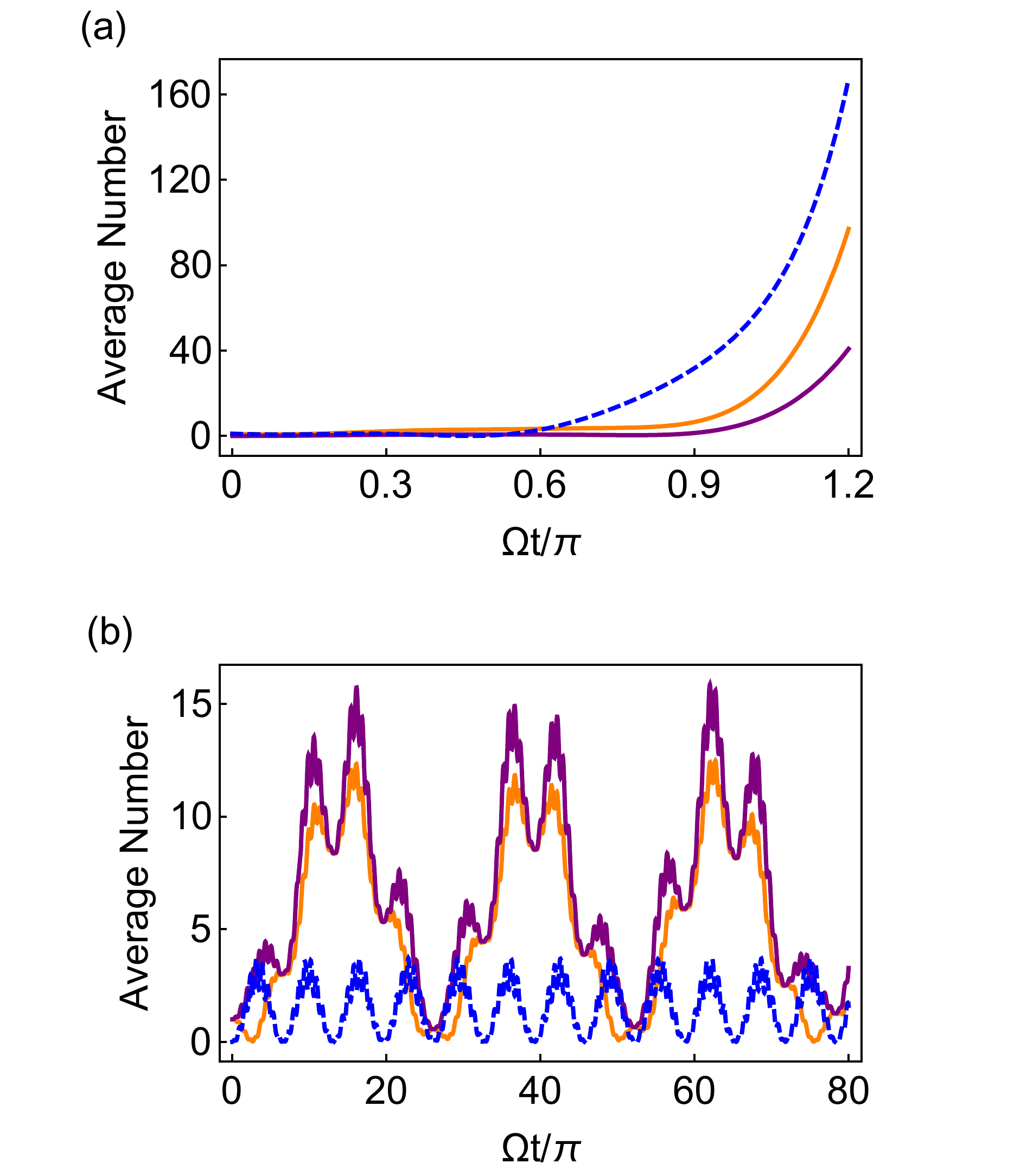}
\end{center}
\caption{Time evolution of average numbers,  $\langle \delta \hat{a}^\dag_1 \delta \hat{a}_1 \rangle$ of the cavity mode 1 (orange solid line), $\langle \delta \hat{a}^\dag_2 \delta \hat{a}_2 \rangle$ of the cavity mode 2 (purple solid line), and $\langle \delta\hat{b}^\dag \delta\hat{b} \rangle$ of the mechanical mode (blue dashed line) in a three-mode optomechanical system with (a) $\Delta_1=\Delta_2=1.5\,\Omega$, $|\kappa_1|=|\kappa_2|/0.6=\Omega$, (b) $\Delta_1=-\Delta_2=1.5\,\Omega$, $|\kappa_1|=3|\kappa_2|/5=0.15\,\Omega$.}
\label{fig:5}
\end{figure}

Similarly to the two-mode system, the system can become unstable even when both cavities are driven by red-detuned lasers ($s=+$, $\epsilon\Delta>0$, $\kappa_s>K_R$), while it can be stable when both cavities are driven by blue-detuned lasers ($s=+$, $\epsilon\Delta<0$, $\kappa_s<K_B$). Besides, the system can be stabilized when $s=-$ and $\epsilon\Delta<0$, which corresponds to one cavity driven by a red-detuned laser while the other is driven by a blue-detuned one and $|\kappa_\text{red}|<|\kappa_\text{blue}|$. Here, $\kappa_\text{red (blue)}$ represents the interaction coupling strength between the cavity mode driven by a red-detuned (blue-detuned) laser and the mechanical mode. This fact implies that the power of the red-detuned laser stronger than that of the blue-detuned one is no longer a necessary condition for stability as the system goes beyond RWA. To illustrate these interesting cases, Fig.~\ref{fig:5} demonstrates the time evolution of average numbers for deviation  modes $\delta\hat{a}_1$, $\delta\hat{a}_2$, and $\delta\hat{b}$ of a three-mode system with $\Delta_1=\pm\Delta_2$. 

\begin{figure*}[t]
\begin{center}
\includegraphics[height=0.27\textwidth]{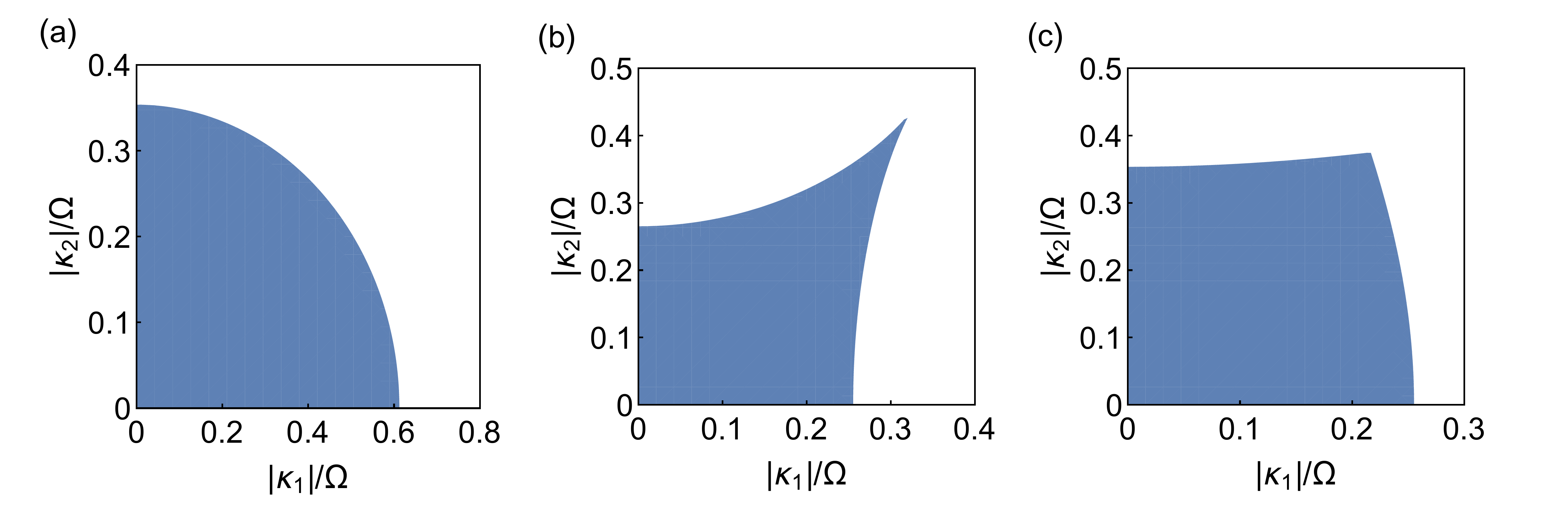}
\end{center}
\caption{Stability diagram of a three-mode optomechanical system as a funtion of $|\kappa_1|/\Omega$ and $|\kappa_2|/\Omega$ when (a) both cavities are red-detuned with $\Delta_1=3\,\Delta_2=1.5\,\Omega$, (b) both cavities are blue-detuned with $\Delta_1=3\,\Delta_2=-1.5\,\Omega$, and (c) cavity 1 is blue-detuned with $\Delta_1=-1.5\,\Omega$, cavity 2 is red-detuned with $\Delta_2=0.5\,\Omega$. Blue area: circular Hamiltonian (dynamically stable), white area: hyperbolic Hamitonian (dynamically unstable).}
\label{fig:6}
\end{figure*}

In addition, the transition of stability from stable to unstable as the system parameters cross the critical boundaries is observed in a three-mode optomechanical system. For $\Delta_1=\pm\Delta_2$, the critical points are the same as ones of a two-mode system with $\Delta$ and $|\kappa|$ replaced by $\epsilon\Delta$ and $\kappa_s$, respectively. For a general system with $\Delta_1$ and $\Delta_2$ arbitrary, the stability phase transition is shown by stability diagrams in Fig. \ref{fig:6}, where the critical boundaries between stable and unstable areas appears in all three situations: (a) both cavities are red-detuned, (b) both cavities are blue-detuned, and (c) one cavity is red-detuned while the other is blue-detuned.

\section{Remarks}\label{con}
We present a novel geometric approach for analyzing the stability of a quantum quadratic system, which serves as a fundamental component in stability theory. Our investigation reveals that a time-independent quantum quadratic system is dynamically unstable if and only if its (diagonalized) Hamiltonian is geometrically hyperbolic. Applying this geometric method to optomechanical systems, we derive comprehensive stability criteria that remain valid across the entire range of system parameters for both two-mode and three-mode configurations. Our findings demonstrate that the system transits its phase from stable to unstable as the system parameters cross the critical boundaries. As a result, the system can be controlled stable in all regimes, surpassing the previous understanding that the system is stabilized by the red-detuning laser for the two-mode case or when the power of the red-detuned laser is stronger than that of the blue-detuned laser for the three-mode case. In addition, we explore the mechanism of the instability in optomechanical systems. An unstable optomechanical system is governed by the hyperbolic Hamiltonian, which originates from the two-mode squeezing interaction term.

\begin{acknowledgments}
This work was supported by
the National Research Foundation of Korea (NRF)
grant funded by the Korea government (MSIT) (No.
2022M3E4A1077369)
\end{acknowledgments}

\appendix
\section{Geometric Hamiltonians of multi-mode quadratic systems}
\label{appA}
In this Appendix, we describe in detail how to rewrite Hamiltonian of a general $N$-mode quadratic system in the geometric form, as in Eq.~\eqref{Geo-H}. We also prove that the geometric Hamiltonian of the system contains at least one hyperbolic modal Hamiltonian when the system is dynamically unstable. 
 
In terms of canonical variables $\hat{\bm{\xi}}=(\hat{x}_1,..., \hat{x}_N, \hat{p}_1,...,$ $\hat{p}_N)^T$, 
the Hamiltonian in Eq.~\eqref{quad-H} is rewritten in the quadratic form of
\begin{equation}\label{quad-Hq}
\hat{H} =\frac{1}{2} \hat{\bm\xi}^T \bm{V} \hat{\bm\xi},
\end{equation}
where $\bm{V}$ is a $2N\times 2N$ real symmetric matrix with constant elements. The Heisenberg equations of motion for $\hat{\bm\xi}$ is given by
\begin{equation}\label{motion-eq}
\frac{d}{dt} \hat{\bm\xi}(t) = \bm{J V}   \hat{\bm\xi}(t) \equiv \bm{A} \hat{\bm\xi}(t),
\end{equation}
where $\bm{J}$ is a skew-symmetric matrix with elements $ J_{ij} = - i [\hat{\xi}_i,\hat{\xi}_j]$, and $\bm{A}=\bm{JV}$ is the equation-of-motion matrix. Because commutation relations $ [\hat{\xi}_i(t),\hat{\xi}_j(t)]=iJ_{ij}$ are preserved regardless of the time evolution, the matrix $\bm{A}$ satisfies the condition
\begin{equation}\label{sol}
\bm{JA} +\bm{A}^{T}\bm{J}=0,
\end{equation}
which restricts eigenvalues of $\bm{A}$  to four classes: nonzero real pairs, purely imaginary pairs, complex quadruplets, and zeros~\cite{LM}. More importantly, matrices satisfying the condition~\eqref{sol} can be transformed into a Jordan normal form $\bm{A}_{J}$ by some symplectic transformation $\bm{S}_J$,
\begin{equation}\label{Tran-to-Jn}
\bm{A}_{J}=\bm{S}_J\bm{A}\bm{S}_J^{-1} = \bm{S}_J \bm{J} \bm{V} \bm{S}_J^{-1}  = \bm{J} (\bm{S}_J^{-1})^T \bm{V} \bm{S}_J^{-1} = \bm{J} \bm{V}_J,
\end{equation}
where $\bm{S}_J$ satisfies the symplectic condition $\bm S_J^T\bm{J}\bm{S}_J=\bm{J}$ and its form varies over classes of eigenvalues of $\bm{A}_J$~\cite{LM}. Under the symplectic transformation $\bm{S}_J$, the canonical variables $\hat{\bm\xi}$ transform to 
\begin{equation}\label{new-var}
\hat{\bm\xi}_J=\bm{S}_J\hat{\bm\xi}.
\end{equation}
The Hamiltonian Eq.~\eqref{quad-Hq} is rewritten in terms of the transformed variables $\hat{\bm\xi}_J$ by
\begin{equation}\label{Jor-H}
\hat{H} = \frac{1}{2}\hat{\bm\xi}^T_J\bm{V}_J\hat{\bm\xi}_J,
\end{equation}
where $\bm{V}_J=\bm{J}^{-1}\bm{A}_J= -\bm{J}\bm{A}_J$. 

Jordan normal forms of $\hat{H}$ are classified in general into 6 types with respect to subsets of eigenvalues of the equation-of-motion matrix $\bm{A}$~\cite{Kus}. In other words, $\bm{V}_J$ (and $\bm{A}_J$) is given in block diagonal form, i.e. 
\begin{equation}\label{block-v}
\bm{V}_J = \bm{V}_{J,1} \oplus \bm{V}_{J,2} \oplus ... = \bigoplus_k \bm{V}_{J, k},
\end{equation}
where the block matrices involve mutually independent subsets of modes, and each block belongs to one of the types. 
Thus it suffices to consider each block matrix separately. Starting with each block matrix $\bm{V}_{J,k}$ (more precisely $\bm{A}_{J, k}$) and identifying its type $\tau$, we employ its Hamiltonian $\hat{H}^\tau_k$ in the normal form, as given in Ref.~\cite{Kus}. We  then show that some symplectic matrix $\bm{S}^\tau_k$ transforms $\hat{H}^\tau_k$ into the form of Eq.~\eqref{dig}, 
\begin{equation} \label{gen-ham}
\hat{H}^\tau_k = \frac{1}{2} \left(\hat{\bm{\Xi}}^\tau_k\right)^T \bm{W}^\tau_k \hat{\bm{\Xi}}^\tau_k = \hat{H}^\tau_{k, G}+\hat{H}^\tau_{k, I},
\end{equation}
where $\hat{\bm{\Xi}}^\tau_k = \bm{S}^\tau_k \hat{\bm{\xi}}_{J,k}^\tau$ and $\bm{W}^\tau_k = \left[\left(\bm{S}^{\tau}_k\right)^{-1}\right]^T\bm{V}^\tau_{J,k}\left(\bm{S}^{\tau}_k\right)^{-1}$. Note that the canonical variables $\hat{\bm{\xi}}^\tau_{J,k}$ and $\hat{\bm{\Xi}}^\tau_k$, and the matrices $\bm{W}^\tau_k$ and $\bm{V}^\tau_{J,k}$, before and after $\bm{S}^\tau_k$, all depend on type $\tau$ and block $k$. However, as separating each block matrix $\bm{V}^\tau_{J, k}$ of type $\tau$, one by one, we omit script set $\{\tau, k, J\}$ in following discussions, in order to simplify notations.

{\em Type I --.} Type I is associated with a pair of  nonzero and real eigenvalues $\{\pm\lambda \vert \lambda > 0\}$ among entire eigenvalues of $\bm{A}$ in Eq.~\eqref{motion-eq}.
Type-I Hamiltonian $\hat{H}$ is given in the Jordan normal form by~(case 1 in Table II, Ref.~\cite{Kus})
\begin{equation} \label{ham-type1}
\hat{H} = \lambda \sum_{j=1}^D \hat{x}_{j} \hat{p}_{j}  + \sum_{j=1}^{D-1} \hat{x}_{j}\hat{p}_{j+1} = \frac{1}{2} \hat{\bm{\xi}}^T \bm{V} \hat{\bm{\xi}},
\end{equation}
where $D$ is the length of Jordan chain for type I and $\bm{V} = \bm{V}_{J,k}$ for the given block matrix $\bm{V}_{J,k}$ of type I in Eq.~\eqref{block-v}. Here, the variables $\hat{\bm{\xi}} = (\hat{x}_1, ..., \hat{x}_D, \hat{p}_1, ..., \hat{p}_D)^T$ of $D$ modes are those of $\hat{\bm{\xi}}_{J, k}$, associated with $\bm{V}_{J, k}$.

We employ the symplectic matrix $\bm{S}$ such that $\bm{S} \hat{\bm{\xi}} = \hat{U} \hat{\bm{\xi}} \hat{U}^\dag$, where
\begin{eqnarray} \label{uni1}
\hat{U} = \exp\left(-i(\pi/8)\sum_{j=1}^D(\hat{p}^2_j+\hat{x}^2_j)\right).
\end{eqnarray}
By $\bm{S}$, $\hat{\bm{\xi}}$ transforms to 
\begin{eqnarray} \label{cap-xi1}
\hat{\bm{\Xi}} :=  (\hat{X}_1, ..., \hat{X}_D, \hat{P}_1, ..., \hat{P}_D)^T = \bm{S} \hat{\bm{\xi}},
\end{eqnarray}
and $\bm{V}$ does to $\bm{W} = \left( \bm{S}^{-1} \right)^T \bm{V} \bm{S}^{-1}$. 
Then, $\hat{H}$ is rewritten by $\hat{H} = \hat{\bm{\Xi}}^T \bm{W} \hat{\bm{\Xi}} / 2= \hat{H}_G + \hat{H}_I$, as in Eq.~\eqref{gen-ham}, where
\begin{eqnarray}
\hat{H}_G&=&\frac{\lambda}{2}\sum_{j=1}^D\big(\hat{P}^2_j-\hat{X}^2_j\big),\\
\hat{H}_I&=&\frac{1}{2}\sum_{j=1}^{D-1}\big(\hat{X}_j\hat{P}_{j+1}-\hat{P}_j\hat{X}_{j+1}-\hat{X}_j\hat{X}_{j+1}+\hat{P}_j\hat{P}_{j+1}\big).\nonumber\\
&&
\end{eqnarray}

{\em Type II --.} Type II is associated with a complex quadruplet of eigenvalues $\{\pm \lambda, \pm \lambda^* \vert \lambda = \lambda_r+i \lambda_i, \lambda_r>0, \lambda_i>0\}$ among entire eigenvalues of $\bm{A}$ in Eq.~\eqref{motion-eq}.
Type-II Hamiltonian $\hat{H}$ is given in the Jordan normal form by~(case 2 in Table II, Ref.~\cite{Kus})
\begin{eqnarray} \label{ham-type2}
\hat{H} &=& \lambda_r \sum_{j=1}^{2D} \hat{x}_j \hat{p}_j + \lambda_i\sum_{j=1}^D\left(\hat{x}_{2j}\hat{p}_{2j-1}-\hat{p}_{2j}\hat{x}_{2j-1}\right)\nonumber\\
&& +\sum_{j=1}^{2D-2} \hat{x}_j \hat{p}_{j+2} = \frac{1}{2} \hat{\bm{\xi}}^T \bm{V} \hat{\bm{\xi}},
\end{eqnarray}
where $D$ is the length of Jordan chain for type II and $\bm{V} = \bm{V}_{J,k}$ for the given block matrix $\bm{V}_{J,k}$ of type II in Eq.~\eqref{block-v}. Here, the variables $\hat{\bm{\xi}} = (\hat{x}_1, ..., \hat{x}_{2D}, \hat{p}_1, ..., \hat{p}_{2D})^T$ of $2D$ modes are those of $\hat{\bm{\xi}}_{J, k}$, associated with $\bm{V}_{J, k}$.

We employ the symplectic matrix $\bm{S}$ such that $\bm{S} \hat{\bm{\xi}} = \hat{U} \hat{\bm{\xi}} \hat{U}^\dag$, where
\begin{equation} \label{uni2}
\hat{U} = \exp\left(-i(\pi/8)\sum_{j=1}^{2D}(\hat{p}^2_j+\hat{x}^2_j)\right).
\end{equation}
By $\bm{S}$, $\hat{\bm{\xi}}$ transforms to 
\begin{equation} \label{cap-xi2}
\hat{\bm{\Xi}} :=  (\hat{X}_1, ..., \hat{X}_{2D}, \hat{P}_1, ..., \hat{P}_{2D})^T = \bm{S} \hat{\bm{\xi}},
\end{equation}
and $\bm{V}$ does to $\bm{W} = \left( \bm{S}^{-1} \right)^T \bm{V} \bm{S}^{-1}$. 
Then, $\hat{H}$ is rewritten by $\hat{H} = \hat{\bm{\Xi}}^T \bm{W} \hat{\bm{\Xi}} / 2= \hat{H}_G + \hat{H}_I$, as in Eq.~\eqref{gen-ham}, where
\begin{eqnarray}
\hat{H}_G&=&\frac{\lambda_r}{2}\sum_{j=1}^{2D}\left(\hat{P}^2_j-\hat{X}^2_j\right),\\
\hat{H}_I&=&\frac{1}{2}\sum_{j=1}^{2D-2}\big(\hat{X}_j\hat{P}_{j+2}-\hat{P}_j\hat{X}_{j+2}-\hat{X}_j\hat{X}_{j+2}+\hat{P}_j\hat{P}_{j+2}\big)\nonumber\\
&&+\lambda_i\sum_{j=1}^{D}\big(\hat{X}_{2j}\hat{P}_{2j-1}-\hat{P}_{2j}\hat{X}_{2j-1}\big).
\end{eqnarray}

{\em Type III --.} Type III is associated with a pair of purely imaginary eigenvalues $\{\pm i \lambda \vert \lambda > 0\}$, with odd $D$, among entire eigenvalues of $\bm{A}$ in Eq.~\eqref{motion-eq}. 
Type-III Hamiltonian $\hat{H}$ is given in the Jordan normal form by~(case 6 in Table II, Ref.~\cite{Kus})
\begin{eqnarray}\label{JnH-Io}
\hat{H} &=&  i\sigma \frac{\lambda}{2}\sum_{j=1}^D(-1)^{j+1}\left(\hat{x}_j\hat{x}_{D+1-j}+\hat{p}_j\hat{p}_{D+1-j}\right) \nonumber\\
&&+ \sum_{j=1}^{D-1}\hat{x}_j\hat{p}_{j+1} = \frac{1}{2} \hat{\bm{\xi}}^T \bm{V} \hat{\bm{\xi}},
\end{eqnarray}
where $D$ is the length of Jordan chain, $i \sigma = \pm 1$ is determined by generating generalized eigenvectors for type III (see Ref.~\cite{Kus}), and $\bm{V} = \bm{V}_{J,k}$ for the given block matrix $\bm{V}_{J,k}$ of type III in Eq.~\eqref{block-v}. Here, the variables $\hat{\bm{\xi}} = (\hat{x}_1, ..., \hat{x}_D, \hat{p}_1, ..., \hat{p}_D)^T$ of $D$ modes are those of $\hat{\bm{\xi}}_{J, k}$, associated with $\bm{V}_{J, k}$.

For $D=1$, the normal form Hamiltonian in Eq.~\eqref{JnH-Io} is itself the circular Hamiltonian of a harmonic oscillator, i.e., $\hat{H} = \hat{H}_G$ with
\begin{equation}
\hat{H}_G =i\sigma \frac{\lambda}{2}\left(\hat{p}_1^2+\hat{x}_1^2\right) \quad \text{and} \quad \hat{H}_I = 0.
\end{equation}

For $D\ge 3$, we employ the symplectic matrix $\bm{S}$ such that $\bm{S} \hat{\bm{\xi}}  = \hat{U}_2\hat{U}_1 \hat{\bm{\xi}} \hat{U}_1^\dag\hat{U}_2^\dag$,
where
\begin{eqnarray} 
\label{uni3a}
\hat{U}_1 &=& \exp\left(i(\hat{x}_1\hat{x}_2-\hat{p}_{D-1}\hat{p}_D)\right), \\
\label{uni3b}
\hat{U}_2 &=& \exp\left(-i(\pi/4)\hat{U}_1(\hat{p}^2_1+\hat{x}^2_1)\hat{U}_1^\dag\right).
\end{eqnarray}
By $\bm{S}$, $\hat{\bm{\xi}}$ transforms to 
\begin{eqnarray} \label{cap-xi3}
\hat{\bm{\Xi}} :=  (\hat{X}_1, ..., \hat{X}_D, \hat{P}_1, ..., \hat{P}_D)^T = \bm{S} \hat{\bm{\xi}},
\end{eqnarray}
and $\bm{V}$ does to $\bm{W} = \left( \bm{S}^{-1} \right)^T \bm{V} \bm{S}^{-1}$. 
Then, $\hat{H}$ is rewritten by $\hat{H} = \hat{\bm{\Xi}}^T \bm{W} \hat{\bm{\Xi}} / 2= \hat{H}_G + \hat{H}_I$, as in Eq.~\eqref{gen-ham}, where
\begin{eqnarray}
\hat{H}_G&=&\hat{P}^2_1+\hat{P}^2_D,\\
\hat{H}_I&=&i\sigma\frac{\lambda}{2}\sum_{j=2}^{D-1}(-1)^{j+1}(\hat{X}_j\hat{X}_{D+1-j}+\hat{P}_j\hat{P}_{D+1-j})+\hat{P}_1\hat{P}_2\nonumber\\
&&+i\sigma\lambda(\hat{P}_1\hat{X}_D-\hat{X}_1\hat{P}_D) + \sum_{j=2}^{D-1}\hat{X}_j\hat{P}_{j+1}.
\end{eqnarray}
We note that $\hat{H}$ involves $D$ modes, even though two modes of quadratures $\hat{P}_{1,D}$ explicitly appear in $\hat{H}_G$.

{\em Type IV --.} Type IV is associated with a pair of purely imaginary eigenvalues $\{\pm i \lambda \vert \lambda > 0\}$, with even $D$, among entire eigenvalues of $\bm{A}$ in Eq.~\eqref{motion-eq}. 
Type-IV Hamiltonian $\hat{H}$ is given in the Jordan normal form by~(case 5 in Table II, Ref.~\cite{Kus})
\begin{eqnarray}
\hat{H} &=&\frac{\sigma}{2}\sum_{j=1}^{D-1} (-1)^{j+1} \left(\hat{x}_j\hat{x}_{D-j}+\hat{p}_{j+1}\hat{p}_{D+1-j}\right)\nonumber\\
&&+\frac{\sigma \lambda}{2}\sum_{j=1}^D\left(\hat{x}_j\hat{x}_{D+1-j}+\hat{p}_j\hat{p}_{D+1-j}\right)=\frac{1}{2} \hat{\bm{\xi}}^T \bm{V} \hat{\bm{\xi}},\nonumber\\  
\end{eqnarray}
where $D$ is the length of Jordan chain, and $\sigma = \pm 1$ is determined by generating generalized eigenvectors for type IV (see Ref.~\cite{Kus}), and $\bm{V} = \bm{V}_{J,k}$ for the given block matrix $\bm{V}_{J,k}$ of type IV in Eq.~\eqref{block-v}. Here, the variables $\hat{\bm{\xi}} = (\hat{x}_1, ..., \hat{x}_D, \hat{p}_1, ..., \hat{p}_D)^T$ of $D$ modes are those of $\hat{\bm{\xi}}_{J, k}$, associated with $\bm{V}_{J, k}$.

For $D=2$, 
we employ the symplectic matrix $\bm{S}$ such that $\bm{S} \hat{\bm{\xi}}  = \hat{U} \hat{\bm{\xi}} \hat{U}^\dag$,
where
\begin{eqnarray} 
\hat{U} &=& \exp\left(-i(\pi/4)(\hat{p}^2_1+\hat{x}^2_1)\right).
\end{eqnarray}
By $\bm{S}$, $\hat{\bm{\xi}}$ transforms to 
\begin{eqnarray} \label{cap-xi4a}
\hat{\bm{\Xi}} :=  (\hat{X}_1, ..., \hat{X}_D, \hat{P}_1, ..., \hat{P}_D)^T = \bm{S} \hat{\bm{\xi}},
\end{eqnarray}
and $\bm{V}$ does to $\bm{W} = \left( \bm{S}^{-1} \right)^T \bm{V} \bm{S}^{-1}$. 
Then, $\hat{H}$ is rewritten by $\hat{H} = \hat{\bm{\Xi}}^T \bm{W} \hat{\bm{\Xi}} / 2= \hat{H}_G + \hat{H}_I$, as in Eq.~\eqref{gen-ham}, where
\begin{equation}
\hat{H}_G=\frac{\sigma}{2}\big(\hat{P}_1^2+\hat{P}_2^2\big), \,  
\hat{H}_I=\sigma\lambda\big(\hat{P}_1\hat{X}_2-\hat{X}_1\hat{P}_2\big).
\end{equation}

For $D\ge 4$, 
we employ the symplectic matrix $\bm{S}$ such that $\bm{S} \hat{\bm{\xi}}  = \hat{U}_2\hat{U}_1 \hat{\bm{\xi}} \hat{U}_1^\dag\hat{U}_2^\dag$,
where
\begin{eqnarray}
\hat{U}_1 &=& \exp\big((-i/2)(\hat{x}_{1}\hat{p}_{D-1}-\hat{x}_{2}\hat{p}_{D})\big), \\
\hat{U}_2 &=& \exp\big(-i(\pi/4)\hat{U}_1(\hat{p}^2_1+\hat{x}^2_1)\hat{U}_1^\dag\big).
\end{eqnarray}
By $\bm{S}$, $\hat{\bm{\xi}}$ transforms to 
\begin{eqnarray} \label{cap-xi4b}
\hat{\bm{\Xi}} :=  (\hat{X}_1, ..., \hat{X}_D, \hat{P}_1, ..., \hat{P}_D)^T = \bm{S} \hat{\bm{\xi}},
\end{eqnarray}
and $\bm{V}$ does to $\bm{W} = \left( \bm{S}^{-1} \right)^T \bm{V} \bm{S}^{-1}$. 
Then, $\hat{H}$ is rewritten by $\hat{H} = \hat{\bm{\Xi}}^T \bm{W} \hat{\bm{\Xi}} / 2= \hat{H}_G + \hat{H}_I$, as in Eq.~\eqref{gen-ham}, where
\begin{eqnarray}
\hat{H}_G&=&\frac{\sigma}{2}\big(\hat{P}^2_1+\hat{P}^2_D\big),\\
\hat{H}_I&=&\sigma\lambda\big(\hat{P}_1\hat{X}_D-\hat{X}_1\hat{P}_D\big)+\sigma\big(\hat{P}_1\hat{X}_{D-1}+\hat{P}_2\hat{P}_D\big)\nonumber\\
&&+ \frac{\sigma}{2}\sum_{j=2}^{D-2} (-1)^{j+1} \big(\hat{X}_j\hat{X}_{D-j}+\hat{P}_{j+1}\hat{P}_{D+1-j}\big)\nonumber\\
&&+\frac{\sigma \lambda}{2}\sum_{j=2}^{D-1}\big(\hat{X}_j\hat{X}_{D+1-j}+\hat{P}_j\hat{P}_{D+1-j}\big).
\end{eqnarray}
We note that $\hat{H}$ involves $D$ modes, even though two modes of quadratures $\hat{P}_{1,D}$ explicitly appear in $\hat{H}_G$.

{\em Type V --.} Type V is associated with eigenvalues of zero $\{\lambda \vert \lambda = 0\}$, with odd $D$, among entire eigenvalues of $\bm{A}$ in Eq.~\eqref{motion-eq}. 
Type-V Hamiltonian $\hat{H}$ is given in the Jordan normal form by~(case 4 in Table II, Ref.~\cite{Kus})
\begin{equation}
\hat{H} =\sum_{j=1}^{D-1}\hat{x}_j\hat{p}_{j+1} = \frac{1}{2} \hat{\bm{\xi}}^T \bm{V} \hat{\bm{\xi}},
\end{equation}
where $D$ is the length of Jordan chain and $\bm{V} = \bm{V}_{J,k}$ for the given block matrix $\bm{V}_{J,k}$ of type V in Eq.~\eqref{block-v}. Here, the variables $\hat{\bm{\xi}} = (\hat{x}_1, ..., \hat{x}_D, \hat{p}_1, ..., \hat{p}_D)^T$ of $D$ modes are those of $\hat{\bm{\xi}}_{J, k}$, associated with $\bm{V}_{J, k}$.

For $D=1$, $\hat{H}=0$ and thus
\begin{equation}
\hat{H}_G = 0 \quad \text{and} \quad \hat{H}_I = 0.
\end{equation}
We note that $\hat{H}$ involves a single mode, even though the Hamiltonians are null.
 
For $D\ge 3$, we employ the symplectic matrix $\bm{S}$ such that $\bm{S} \hat{\bm{\xi}}  = \hat{U}_2\hat{U}_1 \hat{\bm{\xi}} \hat{U}_1^\dag\hat{U}_2^\dag$,
where
\begin{eqnarray} 
\label{uni5a}
\hat{U}_1 &=& \exp\left(i(\hat{x}_1\hat{x}_2-\hat{p}_{D-1}\hat{p}_D)\right), \\
\label{uni5b}
\hat{U}_2 &=& \exp\left(-i(\pi/4)\hat{U}_1(\hat{p}^2_1+\hat{x}^2_1)\hat{U}_1^\dag\right).
\end{eqnarray}
By $\bm{S}$, $\hat{\bm{\xi}}$ transforms to 
\begin{eqnarray} \label{cap-xi5}
\hat{\bm{\Xi}} :=  (\hat{X}_1, ..., \hat{X}_D, \hat{P}_1, ..., \hat{P}_D)^T = \bm{S} \hat{\bm{\xi}},
\end{eqnarray}
and $\bm{V}$ does to $\bm{W} = \left( \bm{S}^{-1} \right)^T \bm{V} \bm{S}^{-1}$. 
Then, $\hat{H}$ is rewritten by $\hat{H} = \hat{\bm{\Xi}}^T \bm{W} \hat{\bm{\Xi}} / 2= \hat{H}_G + \hat{H}_I$, as in Eq.~\eqref{gen-ham}, where
\begin{equation}
\hat{H}_G=\hat{P}^2_1+\hat{P}^2_D \quad \text{and} \quad
\hat{H}_I=\hat{P}_1\hat{P}_2+\sum_{j=2}^{D-1}\hat{X}_j\hat{P}_{j+1}.
\end{equation}
We note that $\hat{H}$ involves $D$ modes, even though two modes of quadratures $\hat{P}_{1,D}$ explicitly appear in $\hat{H}_G$.

{\em Type VI --.} Type VI is associated with eigenvalues of zero $\{\lambda \vert \lambda = 0\}$, with even $D$, among entire eigenvalues of $\bm{A}$ in Eq.~\eqref{motion-eq}. 
Type-VI Hamiltonian $\hat{H}$ is given in the Jordan normal form by~(case 3 in Table II, Ref.~\cite{Kus})
\begin{equation}
\hat{H} = \sigma \sum_{j=1}^{D/2-1} \hat{x}_{j} \hat{p}_{j+1} + \frac{\sigma}{2} (-1)^{D/2+1} \hat{x}^{2}_{D/2} = \frac{1}{2} \hat{\bm{\xi}}^T \bm{V} \hat{\bm{\xi}},
\end{equation}
where $D$ is the length of Jordan chain, and $\sigma = \pm 1$ is determined by generating generalized eigenvectors for type VI (see Ref.~\cite{Kus}), and $\bm{V} = \bm{V}_{J,k}$ for the given block matrix $\bm{V}_{J,k}$ of type VI in Eq.~\eqref{block-v}. Here, the variables $\hat{\bm{\xi}} = (\hat{x}_1, ..., \hat{x}_{D/2}, \hat{p}_1, ..., \hat{p}_{D/2})^T$ of $D/2$ modes are those of $\hat{\bm{\xi}}_{J, k}$, associated with $\bm{V}_{J, k}$.

For $D=2$, we employ the symplectic matrix $\bm{S}$ such that $\bm{S} \hat{\bm{\xi}}  = \hat{U} \hat{\bm{\xi}} \hat{U}^\dag$,
where
\begin{eqnarray} 
\hat{U} &=& \exp\left(-i(\pi/4)(\hat{p}^2_1+\hat{x}^2_1)\right).
\end{eqnarray}
By $\bm{S}$, $\hat{\bm{\xi}}$ transforms to 
\begin{eqnarray} \label{cap-xi6}
\hat{\bm{\Xi}} :=  (\hat{X}_1, ..., \hat{X}_{D/2}, \hat{P}_1, ..., \hat{P}_{D/2})^T = \bm{S} \hat{\bm{\xi}},
\end{eqnarray}
and $\bm{V}$ does to $\bm{W} = \left( \bm{S}^{-1} \right)^T \bm{V} \bm{S}^{-1}$. 
Then, $\hat{H}$ is rewritten by $\hat{H} = \hat{\bm{\Xi}}^T \bm{W} \hat{\bm{\Xi}} / 2= \hat{H}_G + \hat{H}_I$, as in Eq.~\eqref{gen-ham}, where
\begin{equation}
\hat{H}_G = (\sigma/2)\hat{P}_1^2 \quad \text{and} \quad \hat{H}_I = 0.
\end{equation}

For $D\ge 4$, 
we employ the symplectic matrix $\bm{S}$ such that $\bm{S} \hat{\bm{\xi}}  = \hat{U}_2\hat{U}_1 \hat{\bm{\xi}} \hat{U}_1^\dag\hat{U}_2^\dag$,
where
\begin{eqnarray}
\hat{U}_1 &=& \exp\left(i\hat{x}_1\hat{x}_2\right), \\
\hat{U}_2 &=& \exp\big(-i(\pi/4)\hat{U}_1(\hat{p}^2_1+\hat{x}^2_1)\hat{U}_1^\dag
\big).
\end{eqnarray}
By $\bm{S}$, $\hat{\bm{\xi}}$ transforms to 
\begin{eqnarray} \label{cap-xi7}
\hat{\bm{\Xi}} :=  (\hat{X}_1, ..., \hat{X}_{D/2}, \hat{P}_1, ..., \hat{P}_{D/2})^T = \bm{S} \hat{\bm{\xi}},
\end{eqnarray}
and $\bm{V}$ does to $\bm{W} = \left( \bm{S}^{-1} \right)^T \bm{V} \bm{S}^{-1}$. 
Then, $\hat{H}$ is rewritten by $\hat{H} = \hat{\bm{\Xi}}^T \bm{W} \hat{\bm{\Xi}} / 2= \hat{H}_G + \hat{H}_I$, as in Eq.~\eqref{gen-ham}, where
\begin{eqnarray}
\hat{H}_G&=&\sigma\hat{P}^2_1,\\
\hat{H}_I&=&\sigma\hat{P}_1\hat{P}_2+\sigma \sum_{j=2}^{D/2-1} \hat{X}_{j} \hat{P}_{j+1} + \frac{\sigma}{2} (-1)^{D/2+1} \hat{X}^{2}_{D/2}.\nonumber\\
&&
\end{eqnarray}
We note that $\hat{H}$ involves $D/2$ modes, even though a single mode of quadrature $\hat{P}_{1}$ explicitly appear in $\hat{H}_G$.

We note that, for each type $\tau$, $\hat{H}_I^\tau$ commutes with $\hat{H}_G^\tau$. In the interaction picture by the unitary transformation $\hat{U}^{\tau}_I(t)=\exp(-it\hat{H}_I^{\tau})$, the system is governed by the geometric Hamiltonian $\hat{H}^\tau_G(t) = \hat{U}^\tau_I(t) \hat{H}_G^\tau \left( \hat{U}^\tau_I(t) \right)^\dag  = \hat{H}^\tau_G$, which is independent of time $t$, thanks to the commutation $[\hat{H}^\tau_G, \hat{H}^\tau_I] = 0$. 

In general, the quadratic Hamiltonian $\hat{H}$ in Eq.~\eqref{quad-Hq} or Eq.~\eqref{Jor-H} is given by the block diagonal matrix of $N$ modes, $\bm{V}_J = \bigoplus_k \bm{V}_{J, k}$, in other words, $\hat{H} = \sum_k \hat{H}_k$, where $\hat{H}_k = \frac{1}{2} \hat{\bm{\xi}}_{J,k}^T \bm{V}_{J, k} \hat{\bm{\xi}}_{J,k}$ with $\hat{\bm{\xi}}_J = \bigoplus_k \hat{\bm{\xi}}_{J, k}$.  
As $\hat{H}_{k}$'s act mutually exclusive subsets of modes, it holds that $[\hat{H}_k, \hat{H}_l] = 0$ for $k \ne l$. Furthermore, $[H_{k,G}, H_{k,I}]=0$ for each block $k$ and so for each type $\tau$, as shown above. These imply that $[\hat{H}_G, \hat{H}_I] = 0$ for $\hat{H}_G = \sum_k \hat{H}_{k, G}$ and $\hat{H}_I = \sum_k \hat{H}_{k, I}$.
We take the interaction picture by the unitary transformation $\hat{U}_I(t)=\exp(-it\hat{H}_I)$, similarly to the type Hamiltonian $\hat{H}^\tau$. The quadratic system is then generally governed by the geometric Hamiltonian $\hat{H}_G$ from $\hat{H}$ in Eq.~\eqref{quad-Hq}. 

Before closing the Appendix, we discuss the necessary and sufficient conditions for the dynamical instability of quantum system, governed by Hamiltonian in Eq.~\eqref{quad-Hq}, and thereby we elucidate the geometric structure of Hamiltonian when it is unstable. A system is dynamically stable when all solutions of Heisenberg equations in Eq.~\eqref{motion-eq} are bounded for all time $t$~\cite{Nurdin}. This condition is satisfied if and only if the equation-of-motion matrix $\bm{A}$ in Eq.~\eqref{motion-eq} is diagonalizable and has only purely imaginary eigenvalues~\cite{Meyer}, i.e., all eigenvalues of $\bm{A}$ belong to type III with $D=1$, which is the only case that the geometric Hamiltonian contains no hyperbolic (nor lineal) modal Hamiltonian(s). Thus, the system is dynamically unstable if and only if its geometric Hamiltonian contains at least one hyperbolic modal Hamiltonian.

\section{Geometric Hamiltonian of a two-mode optomechanical system}\label{appB}

Here we apply the general results obtained in Appendix \ref{appA} to the two-mode optomechanical system described by Hamiltonian $\hat{H}_{\text{cm-lin}}$ Eq.~\eqref{H2m-q} in order to derive the geometric Hamiltonian listed in Table~\ref{tab:1}. 

(a) For $\Delta > 0$ and $|\kappa| < K_R$, eigenvalues of the equation-of-motion $\bm{A}_\text{cm-lin}$ corresponding to $\hat{H}_\text{cm-lin}$ are two purely imaginary pairs $\pm i\lambda_1$ and $\pm i\lambda_2$, where
\begin{eqnarray}\label{eig-a}
\lambda_1&=&\frac{\sqrt{\Delta^2+\Omega^2+\sqrt{16\Delta\Omega|\kappa|^2+(\Delta^2-\Omega^2)^2}}}{\sqrt{2}},\\
\lambda_2&=&\frac{\sqrt{\Delta^2+\Omega^2-\sqrt{16\Delta\Omega|\kappa|^2+(\Delta^2-\Omega^2)^2}}}{\sqrt{2}},
\end{eqnarray}
with $D_j=1$ and $\sigma_j=-i$ for $j=1,2$. 
The geometric Hamiltonian of mode $j$ ($j=1,2$) for case (a) is then 
\begin{equation}
\hat{H}_j=\frac{\lambda_j}{2}(\hat{P}^2_j+\hat{X}^2_j).
\end{equation}
The symplectic matrix transforming $\hat{x}_j,\hat{p}_j$ to $\hat{X}_j,\hat{P}_j$ is given by
\begin{equation}
\label{Sa}
\bm{S}=\frac{1}{\delta_{12}}
\begin{pmatrix}
\frac{\kappa_i\delta_2}{|\kappa|}\sqrt{\frac{\Delta}{\lambda_{1}}} 
&0 
&-\frac{\kappa_r\delta_2}{|\kappa|}\sqrt{\frac{\Delta}{\lambda_{1}}} 
&-\delta_1\sqrt{\frac{\Omega}{\lambda_1}} \\
-\frac{\kappa_i\delta_2}{|\kappa|}\sqrt{\frac{\Delta}{\lambda_2}} 
&0 
&\frac{\kappa_r\delta_1}{|\kappa|}\sqrt{\frac{\Delta}{\lambda_2}} 
&-\delta_2\sqrt{\frac{\Omega}{\lambda_2}} \\
\frac{\kappa_r\delta_2}{|\kappa|}\sqrt{\frac{\lambda_1}{\Delta}} 
&\delta_1\sqrt{\frac{\lambda_1}{\Omega}} 
&\frac{\kappa_i\delta_2}{|\kappa|}\sqrt{\frac{\lambda_1}{\Delta}} 
&0\\
-\frac{\kappa_r\delta_1}{|\kappa|}\sqrt{\frac{\lambda_2}{\Delta}} 
&\delta_2\sqrt{\frac{\lambda_2}{\Omega}} 
&-\frac{\kappa_i\delta_1}{|\kappa|}\sqrt{\frac{\lambda_2}{\Delta}} &0\\
\end{pmatrix}
\end{equation}
with $\kappa_r:=\text{Re}(\kappa)$, $\kappa_i:=\text{Im}(\kappa)$, $\delta_{12}=\sqrt{\lambda_{1}^2-\lambda_{2}^2}$, and $\delta_j=\sqrt{|\lambda_j^2-\Delta^2|}$ for $j=1,2$.

(b) For $\Delta > 0$ and $|\kappa| = K_R$, matrix $\bm{A}_{\text{cm-lin}}$ has one purely imaginary eigenvalue pair $\pm i\lambda_1=\pm i\sqrt{\Delta^2+\Omega^2}$ with $D_1=1$, $\sigma_1=-i$, and one zero eigenvalue $\lambda_2=0$ with $D_2=2$, $\sigma_2=1$. Then we obtain geometric Hamiltonians
\begin{eqnarray}
\hat{H}_1&=&(\lambda_{1}/2)(\hat{P}_{1}^2+\hat{X}_{1}^2),\\
\hat{H}_2&=&\hat{P}_{2}^2/2,
\end{eqnarray}
and symplectic matrix
\begin{equation}\label{Sb}
\bm{S}=\frac{1}{\Delta\sqrt{\Omega}}
\begin{pmatrix}
\frac{2\Delta\kappa_r}{\sqrt{\lambda_1}}&\frac{\Delta\Omega}{\sqrt{\lambda_1}}&\frac{2\Delta\kappa_i}{\sqrt{\lambda_1}}&0\\
-\frac{2\Omega\kappa_r}{\lambda_1}&\frac{\Delta^2}{\lambda_1}&-\frac{2\Omega\kappa_i}{\lambda_1}&0\\
-\frac{2\Delta^2\kappa_i}{\sqrt{\lambda^3_1}}&0&\frac{2\Delta^2\kappa_r}{\sqrt{\lambda^3_1}}&\frac{\Delta\Omega^2}{\sqrt{\lambda^3_1}}\\
\frac{2\Delta\Omega\kappa_i}{\lambda_1}&0&-\frac{2\Delta\Omega\kappa_r}{\lambda_1}&\frac{\Delta^2\Omega}{\lambda_1}
\end{pmatrix}.
\end{equation}
  
(c) For $\Delta > 0$ and $|\kappa| > K_R$, matrix $\bm{A}_{\text{cm-lin}}$ has one purely imaginary eigenvalue pair $\pm i\lambda_1$ with $D_1=1$, $\sigma_1=-i$, and one real eigenvalue pair $\pm\lambda_2$ with $D_2=1$, where
\begin{eqnarray}
\lambda_1&=&\frac{\sqrt{\Delta^2+\Omega^2+\sqrt{16\Delta\Omega|\kappa|^2+(\Delta^2-\Omega^2)^2}}}{\sqrt{2}},\\
\lambda_2&=&\frac{\sqrt{-\Delta^2-\Omega^2+\sqrt{16\Delta\Omega|\kappa|^2+(\Delta^2-\Omega^2)^2}}}{\sqrt{2}}.
\end{eqnarray}
Hence, we obtain  
\begin{eqnarray}
\hat{H}_1&
=&(\lambda_{1}/2)(\hat{P}_{1}^2+\hat{X}_1^2),\\
\hat{H}_2&=&(\lambda_2/2)(\hat{P}_2^2-\hat{X}_2^2),
\end{eqnarray}
and
\begin{equation}
\label{Sc}
\bm{S}=\frac{1}{s_{12}}
\begin{pmatrix}
\frac{\kappa_r s_2}{|\kappa|}\sqrt{\frac{\lambda_1}{\Delta}} 
&\delta_1\sqrt{\frac{\lambda_1}{\Omega}} 
&\frac{\kappa_i s_2}{|\kappa|}\sqrt{\frac{\lambda_1}{\Delta}} 
&0\\

-\frac{\kappa_r\delta_1}{|\kappa|}\sqrt{\frac{\lambda_2}{\Delta}} 
&s_2\sqrt{\frac{\lambda_2}{\Omega}} 
&-\frac{\kappa_i\delta_1}{|\kappa|}\sqrt{\frac{\lambda_2}{\Delta}} 
&0\\

-\frac{\kappa_i s_2}{|\kappa|}\sqrt{\frac{\Delta}{\lambda_1}} 
&0 
&\frac{\kappa_r s_2}{|\kappa|}\sqrt{\frac{\Delta}{\lambda_1}} 
&\delta_1\sqrt{\frac{\Omega}{\lambda_1}} \\

\frac{\kappa_i \delta_1}{|\kappa|}\sqrt{\frac{\Delta}{\lambda_2}} 
&0 
&-\frac{\kappa_r \delta_1}{|\kappa|}\sqrt{\frac{\Delta}{\lambda_2}} 
&s_2 \sqrt{\frac{\Omega}{\lambda_2}}
\end{pmatrix},
\end{equation}
where $s_{12}=\sqrt{\lambda_1^2+\lambda_2^2}$, $s_2=\sqrt{\lambda_2^2+\Delta^2}$.

(d) For $\Delta=0$, matrix $\bm{A}_{\text{cm-lin}}$ has one purely imaginary eigenvalue pair $\pm i\lambda_1=\pm i\Omega$ with $D_1=1$, $\sigma_1=-i$, and one zero eigenvalue $\lambda_2=0$ with $D_2=2$, $\sigma_2=-1$. Then we obtain
\begin{eqnarray}
\hat{H}_1&=&(\lambda_1/2)(\hat{P}_1^2+\hat{X}_1^2),\\
\hat{H}_2&=&(-1/2)\hat{P}_2^2,
\end{eqnarray}
and
\begin{equation}
\label{Sd}
\bm{S}=
\begin{pmatrix}
\frac{2\kappa_r}{\Omega} & 1 & \frac{2\kappa_i}{\Omega} & 0\\
0 & 0 & \frac{\sqrt{\Omega}}{2\kappa_r} & \frac{-1}{\sqrt{\Omega}}\\
0 & 0 & 0 & 1\\
-\frac{2\kappa_r}{\sqrt{\Omega}} & 0 & -\frac{2\kappa_i}{\sqrt{\Omega}} & 0
\end{pmatrix}.
\end{equation}
 
(e) For $\Delta < 0$ and $|\kappa|<K_B$, matrix $\bm{A}_{\text{cm-lin}}$ has two purely imaginary eigenvalue pairs $\pm i\lambda_1$ and $\pm i\lambda_2$, where
\begin{eqnarray}\label{eig-d}
\lambda_1&=&\frac{\sqrt{\Delta^2+\Omega^2+\sqrt{16\Omega\Delta|\kappa|^2+(\Delta^2-\Omega^2)^2}}}{\sqrt{2}},\\
\lambda_2&=&\frac{\sqrt{\Delta^2+\Omega^2-\sqrt{16\Omega\Delta|\kappa|^2+(\Delta^2-\Omega^2)^2}}}{\sqrt{2}},
\end{eqnarray}
with $D_1=D_2=1$ and $\sigma_1=-\sigma_2=-i\sigma$ with $\sigma=\text{sgn}[\Omega^2-\Delta^2]$. Hence, we obtain
\begin{eqnarray}
\hat{H}_1&=&\sigma(\lambda_1/2)(\hat{P}^2_1+\hat{X}^2_1),\\
\hat{H}_2&=&-\sigma(\lambda_2/2)(\hat{P}^2_2+\hat{X}^2_2),
\end{eqnarray}
and
\begin{equation}
\label{Se}
\bm{S}=\frac{\sqrt{\Omega}}{\delta_{12}}
\begin{pmatrix}
\frac{2\kappa_i\Delta}{\delta_1\sqrt{\lambda_1}}
&0
&-\frac{2\kappa_r\Delta}{\delta_1\sqrt{\lambda_1}} 
&- \frac{\sigma\delta_1}{\sqrt{\lambda_1}}\\

\-\frac{2\kappa_i\Delta}{\delta_2\sqrt{\lambda_2}}
&0
&+ \frac{\kappa_r\Delta}{\delta_2\sqrt{\lambda_2}} 
&\frac{\sigma\delta_1}{\sqrt{\lambda_2}}\\

\frac{2\sigma\kappa_r\sqrt{\lambda_1}}{\delta_1}
&\frac{\delta_1\sqrt{\lambda_1}}{\Omega}
&\frac{2\sigma\kappa_i\sqrt{\lambda_1}}{\delta_1}
&0 \\

\frac{2\sigma\kappa_r\sqrt{\lambda_2}}{\delta_2}
& \frac{\delta_2\sqrt{\lambda_2}}{\Omega}
&\frac{2\sigma\kappa_i\sqrt{\lambda_2}}{\delta_2}
&0 
\end{pmatrix},
\end{equation}
where $\delta_{12}=\sqrt{\lambda_{1}^2-\lambda_{2}^2}$, and $\delta_j=\sqrt{|\lambda_j^2-\Delta^2|}$ for $j=1,2$.

(f) For $\Delta < 0$ and $|\kappa|=K_B$, matrix $\bm{A}_{\text{cm-lin}}$ has one purely imaginary eigenvalue pair $\pm i\lambda=\pm i\sqrt{(\Delta^2+\Omega^2)/2}$ with $D=2$ and $\sigma=\text{sgn}[\Omega^2-\Delta^2]$. Then we obtain
\begin{equation}
\hat{H}_j=(\sigma/2)\hat{P}^2_j,\quad\text{for }j=1,2\\
\end{equation} 
and
\begin{widetext}
\begin{equation}
\label{Sf}
\bm{S}=\sqrt{\frac{\Omega}{2}}
\begin{pmatrix}
\frac{-\kappa_r(\Delta^2+\lambda^2)}{\lambda\sqrt{|\Delta^2-\lambda^2|^3}}& \frac{\sigma(3\lambda^2-\Delta^2)}{2\lambda\Omega\sqrt{|\Delta^2-\lambda^2|}}& \frac{-\kappa_i(\Delta^2+\lambda^2)}{\lambda\sqrt{|\Delta^2-\lambda^2|^3}}& 0\\

\frac{\sigma\kappa_i\Delta(\Delta^2-3\lambda^2)}{\lambda^2\sqrt{|\Delta^2-\lambda^2|^3}}& 0& \frac{\sigma\kappa_r\Delta(3\lambda^2-\Delta^2)}{\lambda^2\sqrt{|\Delta^2-\lambda^2|^3}}& \frac{-(\Delta^2+\lambda^2)}{2\lambda^2\sqrt{|\Delta^2-\lambda^2|}}\\

\frac{-2\kappa_i\Delta}{\lambda\sqrt{|\Delta^2-\lambda^2|}}& 0& \frac{2\kappa_r\Delta}{\lambda\sqrt{|\Delta^2-\lambda^2|}}& \frac{\sigma\sqrt{|\Delta^2-\lambda^2|}}{\lambda}\\

\frac{2\sigma\kappa_r}{\sqrt{|\Delta^2-\lambda^2|}}& \frac{\sqrt{|\Delta^2-\lambda^2}|}{\Omega}& \frac{2\sigma\kappa_i}{\sqrt{|\Delta^2-\lambda^2|}}& 0
\end{pmatrix}.
\end{equation}
\end{widetext}

(g) For $\Delta < 0$ and $|\kappa|>K_B$, eigenvalues of $\bm{A}_{\text{cm-lin}}$ are one complex quadruplet $\pm(\lambda_{r}\pm i\lambda_{i})$, where
\begin{eqnarray}
\lambda_r&=& \frac{\sqrt{-\Delta^2-\Omega^2+\sqrt{4\Delta^2\Omega^2-16\Delta\Omega|\kappa|^2}}}{2},\\
\lambda_i&=& \frac{\sqrt{\Delta^2+\Omega^2+\sqrt{4\Delta^2\Omega^2-16\Delta\Omega|\kappa|^2}}}{2},
\end{eqnarray}
with $D=1$. Then we obtain
\begin{equation}
\hat{H}_j=(\lambda_r/2)(\hat{P}^2_j-\hat{X}^2_j),\quad\text{for } j=1,2
\end{equation}  
and
\begin{widetext}
\begin{equation}\label{Sg}
\bm{S}=\frac{\Lambda}{\Gamma}
\begin{pmatrix}
\frac{\kappa_r(\Delta^2+\Sigma^2)}{4|\kappa|\Delta\Sigma\lambda_i}
& \frac{|\kappa|\Lambda\Sigma +\Sigma^2\lambda_i}{2\Lambda\Sigma^2\lambda_i}
& \frac{\kappa_i(\Delta^2+\Sigma^2)}{4|\kappa|\Delta\Sigma\lambda_i}
& 0\\

-\frac{\kappa_i(|\kappa|\Lambda+\Sigma\lambda_i)}{2|\kappa|\Sigma^2\lambda_i}
&0
&  \frac{\kappa_r(|\kappa|\Lambda+\Sigma\lambda_i)}{2|\kappa|\Sigma^2\lambda_i}
& -\frac{\Lambda(\Delta^2+\Sigma^2)}{4\Delta\Sigma^2\lambda_i}\\

\frac{\kappa_i\left(4|\kappa|\Lambda\lambda_i-\Sigma(\Delta^2-\Omega^2)\right)}{8|\kappa|\Sigma^2\lambda_r\lambda_i}
&0
&\frac{\kappa_r\left(\Sigma(\Delta^2-\Omega^2)-4|\kappa|\Lambda\lambda_i\right)}{8|\kappa|\Sigma^2\lambda_r\lambda_i}
&\frac{\Lambda\left(\lambda_i(\Delta^2-\Sigma^2)-2|\kappa|\Lambda\Sigma\right)}{4\Delta\Sigma^2\lambda_r\lambda_i}\\

\frac{\kappa_r\left(2|\kappa|\Lambda\Sigma-\lambda_i(\Delta^2-\Sigma^2)\right)}{4|\kappa|\Delta\Sigma\lambda_r\lambda_i}
& \frac{\Sigma^2(\Delta^2-\Omega^2)-4|\kappa|\Lambda\Sigma\lambda_i}{8\Lambda\Sigma^2\lambda_r\lambda_i}
&  \frac{\kappa_i\left(2|\kappa|\Lambda\Sigma-\lambda_i(\Delta^2-\Sigma^2)\right)}{4|\kappa|\Delta\Sigma\lambda_r\lambda_i}
&0
\end{pmatrix},
\end{equation}
\end{widetext}
where
\begin{eqnarray*}
\Lambda &=& \sqrt{-\Delta\Omega},\\
\Sigma &=& \sqrt{\lambda_r^2+\lambda_i^2},\\
\Gamma &=& \sqrt{\frac{\Omega}{4\Sigma^2}\left(\frac{\Sigma^2-\Delta^2}{\lambda_r}+\sqrt{\frac{(\Sigma^2-\Delta^2)^2}{\lambda_r^2}+\frac{(\Sigma^2+\Delta^2)^2}{\lambda_i^2}}\right)}.
\end{eqnarray*}

\section{Stability condition of a three-mode optomechanical system}\label{appC}
\begin{table*}
\begin{center}
\caption{Geometric kinds of modal Hamiltonians $\hat{H}_{1,2,3}$ and stability for a three-mode optomechanical system. C: circular, H: hyperbolic (including lineal and zero modes).}
 \label{tab:4}
 \begin{ruledtabular}
\begin{tabular}{lcccc} 
Case   &$\hat{H}_1$ & $\hat{H}_2$ & $\hat{H}_3$ & Stability \\ 
\hline
$\nu^2<\mu^3$	&&&\\
(1) $c_1<0$, $c_2<0$, $c_3<0$	 	& C	& C & C & stable	\\
(2) $c_j<0$, $c_k< 0$, $c_l\geq 0$ for $j,k,l\in\{1,2,3\}$, $j\neq k\neq l$   & C & C & H & unstable\\ 
(3) $c_j<0$, $c_k\geq 0$, $c_l\geq 0$ for $j,k,l\in\{1,2,3\}$, $j\neq k\neq l$   & C & H & H &unstable \\ 
(4) $c_j\geq0$, $c_k\geq 0$, $c_l\geq 0$ for $j,k,l\in\{1,2,3\}$, $j\neq k\neq l$   & H & H & H & unstable\\ 
\hline
$\nu^2=\mu^3$	&	&	& & \\
(5) $4\nu<(\eta_1+\eta_2+3)^3$ & C & H & H & unstable\\
(6) $4\nu\geq(\eta_1+\eta_2+3)^3$ & H & H & H & unstable\\
\hline
$\nu^2>\mu^3$	&	&	& &\\
(7) $\mu+(\nu+\sqrt{\nu^2-\mu^3})^{2/3}<[2(\nu+\sqrt{\nu^2-\mu^3})]^{1/3}(\eta_1+\eta_2+3)$	& C	&	H	&	H & unstable\\
(8) $\mu+(\nu+\sqrt{\nu^2-\mu^3})^{2/3}\geq[2(\nu+\sqrt{\nu^2-\mu^3})]^{1/3}(\eta_1+\eta_2+3)$	& H	&	H	&	H & unstable\\
\end{tabular}
\end{ruledtabular}
\end{center}
\end{table*}

Hamiltonian $\hat{H}_\text{cmc-lin}$ in terms of quadratures has the form
\begin{eqnarray}
\hat{H}_\text{cmc-lin}
&=&\frac{1}{2}\sum_{j=1}^2\Delta_j(\hat{p}^2_j+\hat{x}^2_j)+\frac{1}{2}\sum_{j=1}^2\Omega(\hat{p}^2_b+\hat{x}^2_b)\nonumber\\
&&+2\sum_{j=1}^2(\kappa_{j_r}\hat{x}_j+\kappa_{j_i}\hat{p}_j)(\hat{x}_b+\hat{p}_b),
\end{eqnarray} 
where $\kappa_{j_r}=\text{Re}(\kappa_j)$ and $\kappa_{j_i}=\text{Im}(\kappa_j)$. 
The equation-of-motion matrix corresponding to $\hat{H}_\text{cmc-lin}$ in the basis $\hat{\bm\xi}=(\hat{x}_1,\hat{x}_2,\hat{x}_3,\hat{p}_1,\hat{p}_2,\hat{p}_3)^T$ is given by
\begin{equation}
\bm{A}=
\begin{pmatrix}
0& 0 & 2\kappa_{1i} &\Delta_1 & 0 &0\\
0& 0 & 2\kappa_{2i} & 0 & \Delta_2 & 0\\
0 & 0 & 0 & 0 & 0 & \Omega\\
-\Delta_1 & 0 & -2\kappa_{1r}& 0 & 0 & 0\\
0& -\Delta_2 & -2\kappa_{2r} & 0 & 0 & 0\\
-2\kappa_{1r} & -2\kappa_{2r} &-\Omega & -2\kappa_{1i}& -2\kappa_{2i}& 0 
\end{pmatrix}.
\end{equation}
Matrix $\bm{A}$ has, in general, three eigenvalue pairs ($\pm\sqrt{\lambda_1}$, $\pm\sqrt{\lambda_2}$, $\pm\sqrt{\lambda_3}$) satisfying
\begin{widetext}
\begin{eqnarray}
\lambda_1&=&-\frac{\Omega^2}{3}\left\{3+\eta_1+\eta_2-\frac{\mu+(\nu+\sqrt{\nu^2-\mu^3})^{2/3}}{\left[2(\nu+\sqrt{\nu^2-\mu^3})\right]^{1/3}}\right\},\\
\lambda_2&=&-\frac{\Omega^2}{3}\left\{3+\eta_1+\eta_2+\frac{(1+i\sqrt{3})\mu+(1-i\sqrt{3})\left(\nu+\sqrt{\nu^2-\mu^3}\right)^{2/3}}{2\left[2(\nu+\sqrt{\nu^2-\mu^3})\right]^{1/3}}\right\},\\
\lambda_3&=&-\frac{\Omega^2}{3}\left\{3+\eta_1+\eta_2+\frac{(1-i\sqrt{3})\mu+(1+i\sqrt{3})\left(\nu+\sqrt{\nu^2-\mu^3}\right)^{2/3}}{2\left[2(\nu+\sqrt{\nu^2-\mu^3})\right]^{1/3}}\right\},
\end{eqnarray}
\end{widetext}
where
\begin{eqnarray}
\eta_1&=&\frac{2\Delta^2_1-\Delta^2_2-\Omega^2}{\Omega^2},\qquad\eta_2=\frac{2\Delta^2_2-\Delta^2_1-\Omega^2}{\Omega^2},\\
\mu&=&\frac{2^{\frac{2}{3}}}{3}\bigg(\eta_1^2+\eta_2^2+\eta_1\eta_2+36\frac{\Delta_1|\kappa_1|^2+\Delta_2|\kappa_2|^2}{\Omega^3}\bigg),\\
\nu&=&\eta_1\eta_2(\eta_1+\eta_2)+36\bigg(\frac{\eta_2\Delta_1|\kappa_1|^2}{\Omega^3}+\frac{\eta_1\Delta_2|\kappa_2|^2}{\Omega^3}\bigg).\qquad
\end{eqnarray}
According to Appendix \ref{appA}, the condition under which the geometric Hamiltonian is in the circular form (so that the system is stable) is that  
\begin{equation}
\nu^2-\mu^3<0,\quad c_j<0\quad\text{for all }j=1,2,3, 
\end{equation}
where $c_j=2^{2/3}\sqrt{\mu}\cos(\varphi_j/3)-\eta_1-\eta_2-3$ with $\varphi_j=\arccos(\nu/\sqrt{\mu^3})-2\pi (j-1)$. Otherwise, the geometric Hamiltonian is hyperbolic, and hence the system is unstable. Geometric kinds of modal Hamiltonian $\hat{H}_{1,2,3}$ are represented in Table~\ref{tab:4}.


\begin{thebibliography}{99}
\bibitem{App} M. Aspelmeyer, P. Meystre, and K. Schwab, Phys. Today { \bf 65}, 29 (2012).

\bibitem{MTK} M. Aspelmeyer, T. J. Kippenberg,  and F. Marquardt, Rev. Mod. Phys. {\bf 86}, 1391 (2014).

\bibitem{Rev2} A. K. Sarma, S. Chakraborty, and S. Kalita, AVS Quantum Sci. {\bf 3}, 015901 (2021).


\bibitem{Braginsky} V. B. Braginsky and A. B. Manukin, Sov. Phys. JETP { \bf 25},
653 (1967); V. B. Braginsky, A. B. Manukin, and M. Y. Tikhonov,  ibid. {\bf 31}, 829 (1970).

\bibitem{Dorsel} A. Dorsel, J. D. McCullen, P. Meystre, E. Vignes, and H. Walther, Phys. Rev. Lett. { \bf 51}, 1550 (1983).

\bibitem{Squeezing} C. Fabre, M. Pinard, S. Bourzeix, A. Heidmann, E. Giacobino, and S. Reynaud, Phys. Rev. A { \bf 49}, 1337 (1994); S. Mancini and P. Tombesi, ibid. {\bf 49}, 4055 (1994); X. Y. Lu, Y. Wu, J. R. Johansson, H. Jing, J. Zhang, and F. Nori, Phys. Rev. Lett. {\bf 114}, 093602 (2015).

\bibitem{Cooling} S. Mancini, D. Vitali, and P. Tombesi, Phys. Rev. Lett. { \bf 80}, 688 (1998); P. F. Cohadon, A. Heidmann, and M. Pinard, ibid. { \bf 83}, 3174 (1999); S. Gigan, H. R. Bohm, M. Paternostro, F. Blaser, G. Langer, J. B. Hertzberg, K.C. Schwab, D. Bäuerle, M. Aspelmeyer, and A. Zeilinger, Nature (London) { \bf 444}, 67 (2006); C. Genes, D. Vitali, P. Tombesi, S. Gigan, and M. Aspelmeyer, Phys. Rev. A {\bf 77}, 033804 (2008); J. D. Teufel, T. Donner, D. Li, J. W. Harlow, M. S. Allman, K. Cicak, A. J. Sirois, J. D. Whittaker, K. W. Lehnert, and R. W. Simmonds, ibid. { \bf 475}, 359 (2011).

\bibitem{Entanglement} M. Paternostro, D. Vitali, S. Gigan, M. S. Kim, C. Brukner, J. Eisert, and M. Aspelmeyer, Phys. Rev. Lett. {\bf 99}, 250401(2007); D. Vitali, S. Gigan, A. Ferreira, H. R. B\"ohm, P. Tombesi, A. Guerreiro, V. Vedral, A. Zeilinger, and M. Aspelmeyer, ibid. {\bf 98}, 030405 (2007); R. Ghobadi, A. R. Bahrampour, and C. Simon, Phys. Rev. A {\bf 84}, 033846 (2011); Y. D. Wang and A. A. Clerk, Phys. Rev. Lett. {\bf 110}, 253601 (2013).


\bibitem{OMIT} S. Weis, R. Rivière, S. Delglise, E. Gavartin, O. Arcizet, A. Schliesser, and T. J. Kippenberg, Science {\bf 330}, 1520 (2010); A. H. Safavi-Naeini, T. P. M. Alegre, J. Chan, M. Eichenfield, M. Winger, Q. Lin, J. T. Hill, D. E. Chang, and O. Painter, Nature {\bf 472}, 69 (2011). 

\bibitem{Transducer} K. Stannigel, P. Rabl, A. S. Sørensen, P. Zoller, and M. D. Lukin, Phys. Rev. Lett. { \bf 105}, 220501 (2010); L. Tian, ibid. { \bf 108}, 153604 (2012); L. Tian, Ann. Phys. (Berlin) { \bf 527}, 1 (2015). 

\bibitem{Stability} A. F. Pace and M. J. Collett, Phys. Rev. A {\bf 47}, 3173 (1993).


\bibitem{Amplifier} T. Botter, D. W. C. Brooks, N. Brahms, S. Schreppler, and D. M. Stamper-Kurn, Phys. Rev. A { \bf 85}, 013812 (2012).

\bibitem{o3} T. Wang, L. Wang, Y. M. Liu, C. H. Bai, D. Y. Wang, H. F. Wang, and S. Zhang, Opts. Express {\bf 27}, 29581 (2019).

\bibitem{Hur} A. Hurwitz, Math. Ann. {\bf 46}, 273 (1895). 

\bibitem{Rou} E. J. Routh, \emph{A treatise on the stability of a given state of motion} (Macmillima, London, 1877).

\bibitem{RH} E. X. DeJesus and C. Kaufman, Phys. Rev. A {\bf 35}, 5288 (1987).

\bibitem{Meyer} K. R. Meyer and D. C. Offin, \emph{Introduction to Hamiltonian Dynamical Systems and the N-Body Problem} (Springer International Publishing, Cham, Switzerland, 2017).

\bibitem{Kus} K. Kustura, C. C. Rusconi, and O. R. Isart, Phys. Rev. A {\bf 99}, 022130 (2019).

\bibitem{Jose} J. V. Jose and E. J. Saletan, \emph{Classical Dynamics: A Contemporary Approach} (Cambridge University
Press, Cambridge, 1998).

\bibitem{Arv} Arvind, B. Dutta, N. Mukunda, and R. Simon, Pramana J. Phys. {\bf 45}, 471 (1995).

\bibitem{LM} A. J. Laub and K. Meyer, Celest. Mech. { \bf 9}, 213 (1974).



\bibitem{Knight} C. Gerry and P. Knight, \emph{Introductory Quantum Optics} (The Edinburgh Building, Cambridge, 2005), p. 308.

\bibitem{QPT}  M. J. Hwang, R. Puebla, and M. B. Plenio, Phys. Rev. Lett. {\bf 115}, 180404 (2015).

\bibitem{Bistable}  P. Meystre, E. Wright, J. McCullen, and E.Vignes, J. Opt. Soc. Am. B {\bf 2}, 1830 (1985); A. Gozzini, F. Maccarone one, F.  Mango, I. Longo, and S. Barbarino, ibid. {\bf 2}, 1841 (1985).

\bibitem{Meyer2} K. R. Meyer, J. S. Palacian, P. Yanguas, Discrete Contin. Dynam. Syst. \textbf{33}, 1201 (2013).

\bibitem{NEM}R. Krechetnikov and J. E. Marsden, Rev. Mod. Phys. {\bf 79}, 519 (2007).


\bibitem{Nurdin} H. Nurdin and N. Yamamoto, \emph{Linear Dynamical Quantum
Systems: Analysis, Synthesis, and Control, Communications and Control Engineering} (Springer International Publishing, New York, 2017).

\end{thebibliography}
\end{document}